\documentstyle[pre,aps,preprint]{revtex}

\begin{document}
\draft
\tighten

\title{Equilibrium free energy differences from nonequilibrium
       measurements: a master equation approach}
\author{C. Jarzynski}
\address{Theoretical Astrophysics, T-6, MS B288 \\
         Los Alamos National Laboratory \\
         Los Alamos, NM 87545 \\
         {\tt chrisj@t6-serv.lanl.gov}} 
\date{\today}

\maketitle

\begin{abstract}
It has recently been shown that the Helmholtz free 
energy difference between two equilibrium configurations 
of a system may be obtained from an ensemble of 
{\it finite-time} (nonequilibrium) measurements of 
the work performed in switching an external parameter 
of the system.  Here this result is established, 
as an identity, within the master equation formalism.
Examples are discussed and numerical illustrations
provided.
\end{abstract}


\section*{INTRODUCTION}

Consider some finite classical system which 
depends on an external parameter, $\lambda$.
For instance, the system might be a lattice of
coupled classical spins, and $\lambda$ may 
denote the strength of an externally applied
magnetic field;
or, the system might be a gas of particles,
and $\lambda$ a parameter specifying the 
volume of a box enclosing the gas.
Now suppose that, after allowing the system 
to come to equilibrium with a heat reservoir at
temperature $T$, we change, or ``switch'', the external 
parameter, infinitely slowly, from an initial (say, 
$\lambda=0$) to a final ($\lambda=1$) value.
The system will remain in quasi-static equilibrium
with the reservoir throughout the switching
process, and the total {\it work} performed
on the system will equal the Helmholtz free
energy difference between the initial 
and final configurations\cite{ll15}:
\begin{equation}
\label{eq:infty}
W_\infty = \Delta F \equiv F_1-F_0.
\end{equation}
Here $F_\lambda$ denotes the equilibrium free
energy of the system at temperature $T$, for a
fixed value of $\lambda$. 
The subscript on $W$ reminds us that this
result is valid for infinitely slow switching
of the parameter. 

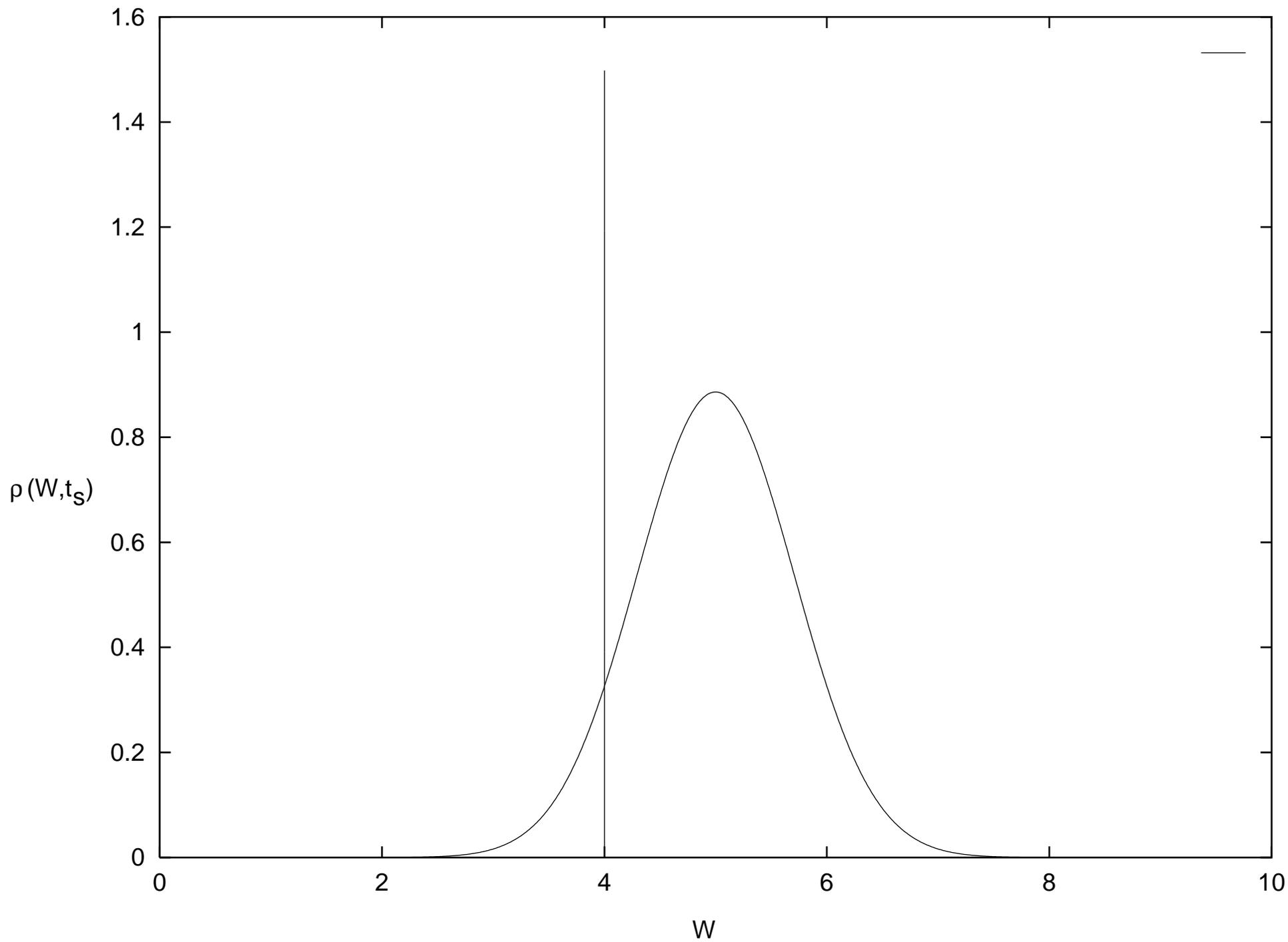
\begin{figure}
\caption{
Distribution of values of work, $\rho(W,t_s)$,
performed during an ensemble of independent
switching measurements at a given
switching time $t_s$.
The vertical line represents a delta
function at $W=\Delta F$, and corresponds
to $t_s\rightarrow\infty$;
in that limit, the work performed during
a single switching process is exactly equal to
$\Delta F$.
The smooth distribution represent $\rho(W,t_s)$
for a finite value of $t_s$.
In this case the ensemble average work
exceeds the free energy difference,
$\overline{W}>\Delta F$, since energy is
dissipated in a finite-time (irreversible)
process.}
\label{fig:schem}
\end{figure}

Now, what happens if, after allowing the system
and reservoir to equilibrate, we switch the value
of $\lambda$ at a {\it finite} rate?
In this case the system will lag behind 
quasi-static equilibrium with the reservoir,
and the total work will depend on the
microscopic initial conditions of system and
reservoir. 
Thus, an {\it ensemble} of such switching measurements 
--- each prepared by first allowing the system to
equilibrate with the reservoir --- 
will yield a distribution of values of $W$. 
Let $\rho(W,t_s)$ denote this distribution, 
where $t_s$ is the ``switching time'' 
over which the value of $\lambda$ is changed from
0 to 1.
(Without loss of generality, assume a uniform
switching rate: $\dot\lambda=t_s^{-1}$.)
In other words, $\rho(W,t_s)\,dW$ is the probability
that the work performed in switching $\lambda$ from
0 to 1, over a time $t_s$, will fall between
$W$ and $W+dW$.
In the limit $t_s\rightarrow\infty$, 
we get $W=\Delta F$ 
(Eq.\ref{eq:infty}), and so  
$\rho\rightarrow\delta(W-\Delta F)$ in this limit.
For finite $t_s$, however:
(1) the distribution $\rho$ acquires a finite
width, reflecting the fluctuations in $W$ from
one switching measurement to the next;
and (2) its centroid shifts to the right:
\begin{equation}
\label{eq:ineq}
\overline{W}\equiv
\int dW\,\rho(W,t_s)\,W\ge\Delta F,
\end{equation}
as a result of dissipation 
(see Fig.\ref{fig:schem})\cite{ll20,wprjeh}.

Eq.\ref{eq:infty} gives the free energy difference 
$\Delta F$ in terms of the work performed during a 
single infinite-time (quasi-static) process.
By contrast, Eq.\ref{eq:ineq} gives
an upper bound on $\Delta F$, from an
ensemble of finite-time (hence, 
nonequilibrium) repetitions of the switching 
process.
Recently\cite{iden}, it was shown that one can
in fact extract the value of $\Delta F$ itself,
not just an upper bound, from the information
contained in $\rho(W,t_s)$.
Specifically, the following result was 
shown to be valid {\it for any switching time} $t_s$:
\begin{equation}
\overline{\exp -\beta W}\equiv
\label{eq:iden}
\int dW\,\rho(W,t_s)\,
\exp -\beta W =
\exp -\beta\Delta F,
\end{equation}
where $\beta^{-1}\equiv k_BT$.
This result gives the value of an equilibrium
quantity, $\Delta F$, in terms of an 
ensemble of finite-time, nonequilibrium measurements.
As discussed in Ref.\cite{iden} and in
Section \ref{sec:comp} below,
the inequality $\overline{W}\ge\Delta F$,
Eq.\ref{eq:ineq}, follows immediately from
Eq.\ref{eq:iden}.

Eq.\ref{eq:iden}
was derived by treating the system of interest
and reservoir, coupled together, as a large,
isolated, classical Hamiltonian system. 
The Hamiltonian governing the motion in the full
phase space was taken to be a sum of three terms:
one for the system of interest ($H_\lambda$), 
one for the reservoir, and a final term, $h_{int}$,
coupling the two.
The magnitude of the interaction term $h_{int}$
was explicitly assumed to be negligible in 
comparison with the other two terms.
With these assumptions, Eq.\ref{eq:iden}
follows from the properties of
Hamiltonian evolution (see Ref.\cite{iden} for details).

The purpose of the present paper is to re-derive
the same result within a different framework.
Instead of considering deterministic evolution
in the full phase space --- 
containing both the system of interest and
reservoir degrees of freedom --- we will treat
only the evolution of the system of interest itself,
described by a trajectory ${\bf z}(t)$.
This evolution will be stochastic rather than
deterministic, and will be governed by a master
equation.
We will assume that this stochastic evolution is
Markovian, and that it satisfies detailed balance.

In presenting this alternative derivation, 
we are motivated by several factors.
First, master equations are a common tool
in statistical physics, therefore it
is reassuring to see that
Eq.\ref{eq:iden} follows in a natural way from
the master equation approach.
Secondly, in this derivation
there is no need to explicitly assume weak
coupling between system and reservoir,
since a reservoir {\it per se}
does not enter into the analysis;
given the assumptions stated below,
Eq.\ref{eq:iden} is {\it identically} true.
Finally, one might come away from Ref.\cite{iden}
with the feeling that that the validity of 
Eq.\ref{eq:iden} depends directly on the properties of
Hamiltonian evolution, in particular Liouville's
theorem.
The treatment herein dispels this
notion: Hamilton's equations appear nowhere
in the derivation.
This point is particularly relevant in the 
context of numerical simulations, where the
evolution of a thermostatted system 
is often realized with the use of non-Hamiltonian
equations of motion.

The plan of this paper is as follows.
In Section \ref{sec:prelim} we establish notation
and terminology, and specify precisely our
assumptions regarding Markovian evolution and
detailed balance.
In Section \ref{sec:deriv} we derive our central
result.
In Section \ref{sec:examp} we consider specific
examples of stochastic processes for which the
result is valid.
In Section \ref{sec:comp} we briefly discuss
the possible utility of Eq.\ref{eq:iden} to the 
numerical computation of free energy differences.
In Section \ref{sec:numres} we illustrate our
central result with numerical simulations, and we
conclude with a few remarks in Section \ref{sec:sum}.
Two appendices provide derivations of results used
in the main text.

\section{PRELIMINARIES}
\label{sec:prelim}

We begin by specifying precisely what we mean by the 
terms {\it work} and {\it free energy} throughout this paper.
We assume that there exists a phase space of
variables (e.g.\ the positions and momenta of constituent
particles), such that the instantaneous microscopic
state of our
system is completely described by the values
of these variables.
Let {\bf z} denote a point in this phase space.
The evolution of our system with time is then described
by a trajectory ${\bf z}(t)$.
The kind of evolution which  will most interest us is
not the evolution of an isolated system, but
rather that of a system in contact with a heat bath.
Hence the trajectory ${\bf z}(t)$ will in general be
{\it stochastic}, reflecting the ``random'' influence
of the heat bath.

Next, assume the existence of a parameter-dependent
Hamiltonian $H_\lambda({\bf z})$.
The Hamiltonian is just a function which, for a fixed
value of $\lambda$, gives the total energy
of the system of interest, in terms of its instantaneous
state ({\bf z}).
The value of $\lambda$ thus parametrizes the external
forces acting on the system
(arising from, e.g., external fields, confining 
potentials, etc.)
When the system is isolated, $H_\lambda$
happens to generate the time evolution of the
system, through Hamilton's equations, but we will
not make use of this property in deriving our central
result.

Given $H_\lambda({\bf z})$, we now define
\begin{mathletters}
\label{eq:fdef}
\begin{eqnarray}
Z_\lambda(\beta) &\equiv& \int d{\bf z}\,
\exp -\beta H_\lambda({\bf z})\\
F_\lambda(\beta) &\equiv& -\beta^{-1}
\ln Z_\lambda(\beta),
\end{eqnarray}
\end{mathletters}
where $\beta$ is a real, positive constant.
(We are being a bit cavalier with units here:
$Z_\lambda$ should be divided by a constant
to make it dimensionless.
However, since this constant only shifts the value 
of $F_\lambda$ by a fixed amount, and therefore
does not affect the free energy difference
$F_1-F_0$, we ignore it.)
$Z_\lambda(\beta)$ and $F_\lambda(\beta)$ 
are of course the {\it partition function}
and {\it free energy} of the system, respectively,
and $\beta^{-1}$ is the temperature, in units
of energy ($\beta\equiv 1/k_B T$).
However, in following the derivation presented
below, it may be most convenient to view these
quantities abstractly, rather than in connection
with their physical significance: $\beta$ is just
some positive constant, and $Z_\lambda(\beta)$ and
$F_\lambda(\beta)$ are the functions of $\beta$
defined by Eq.\ref{eq:fdef}.
In what follows, we will never compare free
energies or partition functions at different
temperatures, hence the dependence of $F_\lambda$
and $Z_\lambda$ on $\beta$ will be left implicit.

The central quantity of interest in this paper will
be the free energy difference
$\Delta F\equiv F_1 - F_0$, for a fixed value
of $\beta$.

As mentioned, the time evolution of the system of
interest is described entirely by a (stochastic) 
phase space trajectory ${\bf z}(t)$.
In general, this evolution will depend on the
externally imposed time-dependence of $\lambda$,
describing the changing external fields to which
the system is subject.
Let us consider the evolution of
the system from an initial time, $t=0$, to a
final time, $t=t_s$, over which the value of
$\lambda$ is ``switched'' from 0 to 1 at a uniform
rate: $\lambda(t) = t/t_s$.
Given this time-dependence of $\lambda$, and given
the trajectory ${\bf z}(t)$ describing the evolution
of the system, the
total {\it work} performed on the system 
is the time integral of $\dot \lambda\,
\partial H_\lambda/\partial\lambda$ along the
trajectory:
\begin{equation}
\label{eq:work}
W\equiv\int_0^{t_s} dt\,\dot\lambda\,
{\partial H_\lambda\over\partial\lambda}
\Bigl({\bf z}(t)\Bigr),
\end{equation}
where $\dot\lambda =d\lambda/dt = t_s^{-1}$.
For the evolution of an isolated
Hamiltonian system, this reduces to
$W=H_1({\bf z}(t_s)) - H_0({\bf z}(0))$, by
Hamilton's equations;
in this case the work performed on the system
is just the change in its energy.
For a system in contact with a heat bath, however,
this no longer holds, since there is a constant
exchange of energy with the bath.

We will assume that the evolution
of our system in phase space is a {\it Markov
process}\cite{kubo}.
This means that the stochastic evolution 
${\bf z}(t)$ is entirely characterized by the
transition probability function
$P({\bf z}^\prime, t\vert {\bf z},t+\Delta t)$.
This gives the probability distribution for 
finding the system in a state {\bf z} at time
$t+\Delta t$, given that at an earlier time $t$
it was known to be at ${\bf z}^\prime$.
Taking the derivative of $P$ with respect to 
$\Delta t$ and evaluating at $\Delta t\rightarrow 0^+$,
we get a function
\begin{equation}
\label{eq:defr}
R({\bf z}^\prime,{\bf z};t)\equiv
{\partial\over\partial(\Delta t)}
P({\bf z}^\prime,t\vert{\bf z},t+\Delta t)
\Biggr\vert_{\Delta t\rightarrow 0^+},
\end{equation}
which gives the instantaneous transition rate
from ${\bf z}^\prime$ to {\bf z}, at time $t$.
The dependence of $R$ on time arises
through whatever external parameters of the
system and reservoir are available and 
time-dependent.
In our case we assume only one such parameter,
$\lambda$ --- characterizing external forces ---
therefore we write
\begin{equation}
R({\bf z}^\prime,{\bf z};t)\rightarrow
R_\lambda({\bf z}^\prime,{\bf z}).
\end{equation}
In other words, the instantaneous transition
rate $R$ from ${\bf z}^\prime$ to {\bf z}
depends on $t$ only through the
value of $\lambda(t)$.

Let us now shift our focus from the description
of a single system, to that of an {\it ensemble}
of systems, each evolving according to the 
stochastic Markov process just described.
This ensemble represents infinitely many
independent realizations of the switching
process.
If $f({\bf z},t)$ denotes the time-dependent
distribution of this ensemble in phase space,
then this distribution obeys
\begin{mathletters}
\label{eq:mast}
\begin{equation}
{\partial f\over\partial t}({\bf z},t)=
\int d{\bf z}^\prime\,f({\bf z}^\prime,t)\,
R_\lambda({\bf z}^\prime,{\bf z}),
\end{equation}
where $\lambda=\lambda(t)$.
We will abbreviate this as
\begin{equation}
{\partial f\over\partial t} = 
\hat R_\lambda\,f,
\end{equation}
\end{mathletters}
where $\hat R_\lambda$ is a linear operator
acting on the space of phase space densities $f$.
Eq.\ref{eq:mast} is our master equation;
when the time-dependence of $\lambda$ is known, 
and an initial distribution $f_0$
is specified, Eq.\ref{eq:mast} uniquely
determines the subsequent evolution of the 
phase space density $f$.

In addition to the Markov assumption
(Eq.\ref{eq:mast}), we will impose another assumption
on our stochastic process: detailed balance.
If $\lambda$ is held fixed, ${\bf z}(t)$ becomes
a {\it stationary} Markov process, describing
the evolution
of a system in contact with a heat reservoir,
when the external forces acting on the system
are time-independent.
Under such evolution, the canonical distribution
in phase space (corresponding to the fixed
value of $\lambda$) ought to be invariant.
By Eq.\ref{eq:mast}, this is equivalent to the statement
that $\hat R_\lambda$ annihilates the 
canonical distribution:
\begin{equation}
\label{eq:stat}
\hat R_\lambda\,\exp -\beta H_\lambda({\bf z})
= 0.
\end{equation}
This places a condition on the linear operator
$\hat R_\lambda$.
We assume that our stochastic process satisfies
this condition, and will refer to this 
assumption as {\it detailed balance}.
(Usually, detailed balance is expressed as the
following, somewhat stronger assumuption:
\begin{equation}
\label{eq:otherdb}
{R_\lambda({\bf z}^\prime,{\bf z}) \over
R_\lambda({\bf z},{\bf z}^\prime)} =
{\exp -\beta H_\lambda({\bf z}) \over
\exp -\beta H_\lambda({\bf z}^\prime)}.
\end{equation}
Eq.\ref{eq:stat} follows immediately from 
Eq.\ref{eq:otherdb}: just set the products of
the cross terms equal, then integrate over 
${\bf z}^\prime$.
The distinction between Eqs.\ref{eq:stat} and
\ref{eq:otherdb} is of little importance in 
the present context, so for simplicity we
refer to Eq.\ref{eq:stat} as detailed balance.)

Having been led to assume detailed balance by
considering the behavior of the system when $\lambda$
is fixed, it may seem natural to make another
assumption.
Namely, if our stochastic process is meant to 
describe a system in contact with a reservoir, then
for fixed $\lambda$ we expect the system to
thermalize:
after an initial relaxation time, the system
samples its phase space canonically;
equivalently, an arbitrary initial ensemble 
$f_0({\bf z})$ relaxes
to a canonical distribution in phase space.
We may express this assumption formally, as:
\begin{equation}
\label{eq:therm}
\lim_{t\rightarrow\infty}
\hat U_\lambda(t) f_0({\bf z}) = 
{1\over Z_\lambda}
\exp -\beta H_\lambda({\bf z}),
\end{equation}
for any normalized $f_0({\bf z})$.
The operator $\hat U_\lambda(t)\equiv\exp \hat R_\lambda t$ 
appearing on the left is just the evolution operator
corresponding to the equation of motion
$\partial f/\partial t = \hat R_\lambda f$, 
for fixed $\lambda$.
Eq.\ref{eq:therm} says that any initial 
distribution $f_0({\bf z})$ relaxes to a 
canonical distribution, and then stays there.
We will refer to this as the assumption of
{\it thermalization}.
Note that while thermalization (Eq.\ref{eq:therm})
implies detailed balance (Eq.\ref{eq:stat}), 
the converse is not true.
Since it will turn out that
the proof of Eq.\ref{eq:iden} requires only the
weaker assumption of detailed balance, and not
the stronger thermalization assumption,
we will assume in what follows that Eq.\ref{eq:stat}
holds, {\it but not necessarily Eq.\ref{eq:therm}},
unless explicitly stated.

[Physically, we expect both thermalization and
detailed balance to hold, as long as our master
equation describes a system in contact with a
genuine heat reservoir.
However, one can easily imagine situations in
which the evolution satisfies Eq.\ref{eq:stat}
but not Eq.\ref{eq:therm}.
A specific example, which we will analyze in
Section \ref{sec:examp}, is that of an isolated
Hamiltonian system.
Note that if thermalization does not hold, then
the validity of Eq.\ref{eq:infty}, $W_\infty=\Delta F$,
is not guaranteed, since that result assumes that
during an infinitely slow switching process 
the system remains in quasi-static equilibrium with
a reservoir.
Eq.\ref{eq:iden} nevertheless remains valid, provided
detailed balance is satisfied.]

At this point we are ready to proceed with a proof
of Eq.\ref{eq:iden}.
Concrete examples of stochastic processes satisfying
the assumptions made in this section will be
discussed in Section \ref{sec:examp}, and 
numerically simulated in Section \ref{sec:numres}.

\section{DERIVATION}
\label{sec:deriv}

The overbar appearing on the left side of 
Eq.\ref{eq:iden} denotes an average over an infinite 
ensemble of independent realizations (repetitions)
of the switching process.
Each realization is described by a trajectory 
${\bf z}(t)$, $0\le t\le t_s$, specifying the
evolution of the complete set of phase space 
variables as the external parameter $\lambda$ is
switched from 0 to 1.
The entire ensemble is then described by a 
time-dependent phase space density:
at any time $t$, $f({\bf z},t)$ represents a
snapshot of the distribution of 
trajectories in phase space.
Since we assume that the system equilibrates 
with the reservoir
(with $\lambda$ held fixed at 0) before the 
start of each realization, we have a canonical
distribution of initial conditions\cite{initial}:
\begin{equation}
\label{eq:initial}
f({\bf z},0) = Z_0^{-1}\,
\exp -\beta H_0({\bf z}).
\end{equation}
During the switching process, however, the
ensemble does not (in general) remain in
instantaneous canonical equilibrium.
In other words, although the distribution
$f=Z_\lambda^{-1}\exp -\beta H_\lambda$
is a (stationary) solution of Eq.\ref{eq:mast}
for $\lambda$ {\it fixed}, it is not, in
general, a solution of Eq.\ref{eq:mast} when
$\lambda$ depends on time.
Thus, for $t>0$, a snapshot of the trajectories
will reveal a distribution $f({\bf z},t)$ 
which lags behind the canonical distribution
corresponding to $\lambda(t)$.
The amount of lag present by the time a given 
value of $\lambda$ is reached, will depend on
how rapidly or slowly we performed the switching
on the way to that value.

For every trajectory ${\bf z}(t)$ in our
ensemble, we can compute the total
work $W$ performed on the system 
(Eq.\ref{eq:work}).
Our task is now to evaluate the ensemble
average of $\exp -\beta W$.
To do this, let us first define, for a given
trajectory ${\bf z}(t)$, a function $w(t)$
which is the ``work accumulated'' up to
time $t$:
\begin{equation}
\label{eq:wacc}
w(t) = \int_0^t dt^\prime\,
\dot\lambda\,
{\partial H_\lambda\over\partial\lambda}
\Bigl({\bf z}(t^\prime)\Bigr).
\end{equation}
Thus, $W = w(t_s)$.
Now consider all those trajectories in the 
ensemble which happen to pass through the
phase space point {\bf z} at time $t$, 
and let $Q({\bf z},t)$ denote the average
value of $\exp -\beta w(t)$, over this particular
subset of trajectories.
Finally, define
\begin{equation}
\label{eq:gdef}
g({\bf z},t) = f({\bf z},t)\,
Q({\bf z},t).
\end{equation}
Note that $g({\bf z},0)=f({\bf z},0)$, 
since $w(0)=0$ for every trajectory.
Given these definitions, the ensemble
average of $\exp -\beta W$ may be expressed as:
\begin{equation}
\label{eq:avex}
\overline{\exp -\beta W} = 
\int d{\bf z}\,g({\bf z},t_s).
\end{equation}
We will now solve for $g({\bf z},t_s)$.

The function $g({\bf z},t)$ obeys 
\begin{equation}
\label{eq:dgdt}
{\partial g\over\partial t} = 
\Biggl(\hat R_\lambda - \beta\dot\lambda
{\partial H_\lambda\over\partial\lambda}\Biggr) g,
\end{equation}
with $\lambda=\lambda(t)$.
To see this, imagine for a moment that each
trajectory ${\bf z}(t)$ in the ensemble represents
a ``particle'' moving about in phase space.
Furthermore, assume that the ``mass'' of each
particle is time-dependent, and is given by
$\mu(t) = \exp -\beta w(t)$.
Each particle thus begins with mass unity:
$\mu(0)=1$.
The function $Q({\bf z},t)$ is then the average
mass of those particles which are found at a
point ${\bf z}$ at time $t$, and $g({\bf z},t)$
represents the time-dependent {\it mass density}
in phase space
(after normalization of the number density to
unity: $\int f\,d{\bf z} = 1$).
This mass density is time-dependent for two
reasons.
First, (A) the mass of each particle changes with
time, at a rate proportional to its instantaneous
mass:
\begin{equation}
\dot\mu(t) = -\beta\dot w(t)\mu(t)
= -\beta\dot\lambda
{\partial H_\lambda\over\partial\lambda}
\Bigl({\bf z}(t)\Bigr)\mu(t).
\end{equation}
This contributes a term
\begin{equation}
{\partial g\over\partial t}\Biggl)_A = 
-\beta\dot\lambda
{\partial H_\lambda\over\partial\lambda}
({\bf z})\,g({\bf z},t)
\end{equation}
to $\partial g/\partial t$.
Secondly, (B) the mass density evolves due to
the flow of particles, described by our master
equation, $\partial f/\partial t=\hat R_\lambda\,f$.
This flow of particles contributes a term
\begin{equation}
{\partial g\over\partial t}\Biggl)_B = 
\hat R_\lambda\,g.
\end{equation}
Adding these contributions, A and B, gives 
Eq.\ref{eq:dgdt}.
An alternative derivation of this evolution equation, 
based on a path integral formulation, is
given in Appendix A.

Given the initial conditions
$g({\bf z},0) = f({\bf z},0) = 
Z_0^{-1}\exp -\beta H_0({\bf z})$,
Eq.\ref{eq:dgdt} is solved by
\begin{equation}
\label{eq:goft}
g({\bf z},t) = Z_0^{-1}
\exp -\beta H_\lambda({\bf z})
= {Z_\lambda\over Z_0} \, f_\lambda^C({\bf z}),
\end{equation}
where $\lambda=\lambda(t)$, 
and $f_\lambda^C$ denotes the canonical distribution
in phase space, for a given value of $\lambda$.
This result is easily verified with the help of
Eq.\ref{eq:stat}.
Then, using Eq.\ref{eq:avex},
we finally arrive at
\begin{equation}
\overline{\exp -\beta W} = 
Z_0^{-1} \int d{\bf z}\,
\exp -\beta H_1({\bf z}) 
= {Z_1\over Z_0} = \exp -\beta\Delta F.
\end{equation}
{\it Q.E.D.}

A different proof of Eq.\ref{eq:iden} has been
discovered by Gavin E.\ Crooks\cite{crooks}.

It is worthwhile to draw attention to
a curious feature of the evolution of
our ``mass density'', $g({\bf z},t)$.
Recall that the evolution of $f({\bf z},t)$
depends nontrivially on the rate at which we switch
$\lambda$:
at finite switching rates, $f({\bf z},t)$ lags
behind the instantaneous equilibrium distribution.
By contrast, the time dependence of 
$g({\bf z},t)$ is very simple:
the mass density is determined uniquely
by the instantaneous value of $\lambda$;
see Eq.\ref{eq:goft}.
Thus, no matter how slowly or rapidly we
switch $\lambda$ from 0 to 1, $g({\bf z},t)$
will always evolve through exactly the
same continuous sequence of ``canonical'' 
mass densities, specified by Eq.\ref{eq:goft};
$g$ {\it never develops a lag}, in the sense that
$f$ does.

Eq.\ref{eq:goft} also implies the result
\begin{equation}
\label{eq:tiden}
\overline{\exp -\beta w(t)} = 
{Z_\lambda\over Z_0} = 
\exp -\beta(F_\lambda - F_0),
\end{equation}
$\lambda=\lambda(t)$, which is in effect a
restatement of our central result, Eq.\ref{eq:iden}.

\section{EXAMPLES}
\label{sec:examp}

We have shown that Eq.\ref{eq:iden} is true for
Markovian processes satisfying detailed balance.
Let us consider a few examples of such processes.

\subsection{Hamiltonian evolution}
\label{subsec:hamev}

As a first case, we take deterministic evolution
under the Hamiltonian $H_\lambda$, as $\lambda$
is varied from 0 to 1:
\begin{equation}
\label{eq:ham}
\dot{\bf z} =
\{{\bf z},H_\lambda\},
\end{equation}
where $\{\cdot,\cdot\}$ denotes the Poisson
bracket.
Ordinarily, one would not call this a stochastic
process, but of course it may be viewed as a 
special case of such.
This evolution is Markovian, and 
$\hat R_\lambda$ is just the Poisson bracket
operator:
\begin{equation}
\hat R_\lambda f = 
\{H_\lambda,f\}.
\end{equation}
It immediately follows that detailed balance
(Eq.\ref{eq:stat}) is satisfied.
The central result of this paper then tells us
that, if we (1) start with a canonical 
distribution of initial conditions, 
$f({\bf z},0) = Z_0^{-1}\exp -\beta H_0({\bf z})$,
and (2) allow trajectories to evolve deterministically
from these initial conditions (Eq.\ref{eq:ham})
as $\lambda$ is varied from 0 to 1, then
this ensemble of trajectories will satisfy
Eq.\ref{eq:iden},
regardless of how slowly or quickly we perform
the switching.
This result was proven more directly in Ref.\cite{iden}.

Since evolution under Eq.\ref{eq:ham}
describes an isolated system (no heat bath),
the thermalization assumption (Eq.\ref{eq:therm}) is
not met, and
the work performed in the limit of infinitely
slow switching will in general not equal
the free energy difference:
$W_\infty\ne\Delta F$.
Eq.\ref{eq:iden} nevertheless remains true,
even in this limit.

This last statement is easily illustrated
with a one-dimensional harmonic oscillator.
Let the Hamiltonian be
\begin{equation}
\label{eq:hoham}
H_\lambda(x,p) = {p^2\over 2} +
\omega_\lambda^2 {x^2\over 2}.
\end{equation}
For a given temperature, the partition
function is given by 
$Z_\lambda = 2\pi/\beta\omega_\lambda$,
and the free energy difference is
\begin{equation}
\Delta F = -\beta^{-1}\ln {Z_1\over Z_0} = 
\beta^{-1}\ln {\omega_1\over\omega_0}.
\end{equation}
Let us now imagine a trajectory ${\bf z}(t)$
evolving under this Hamiltonian, as $\lambda$
is changed infinitely slowly from 0 to 1;
assume for specificity that $\omega_1>\omega_0$.
Since $H_\lambda/\omega_\lambda$ is an
adiabatic invariant for the harmonic 
oscillator\cite{ad_inv}, we have 
$E_1 = (\omega_1/\omega_0)\,E_0$, where
$E_0$ ($E_1$) is the initial (final) energy
of the oscillator.
The work performed on an isolated 
system is equal to the change in its energy, so 
we get $W_\infty = [(\omega_1/\omega_0)-1] E_0$.
A canonical distribution of initial
energies $E_0$ then leads, after some algebra, to the 
following ensemble distribution of values of $W$:
\begin{equation}
\lim_{t_s\rightarrow\infty}
\rho(W,t_s) = 
{\omega_0\beta\over{\omega_1-\omega_0}}
\exp \Biggl(
-{\omega_0\beta W\over\omega_1-\omega_0}
\Biggr)
\theta(W),
\end{equation}
where $\theta$ denotes the unit step function.
It is straighforward to verify that this
distribution satisfies Eq.\ref{eq:iden}.

As another example, consider a single particle bouncing
around inside a three-dimensional cavity with hard
walls, where the shape of the cavity is a function of
$\lambda$.
Let ${\cal V}_\lambda$ denote the 
volume of the cavity.
The free energy difference is then 
$\Delta F=\beta^{-1}\ln({\cal V}_0/{\cal V}_1)$.
Assume that, when $\lambda$ is held fixed, the motion
of the particle is ergodic over the energy
shell (the 5-dimensional surface of constant energy in
6-dimensional phase space); also assume
that ${\cal V}_1\le{\cal V}_0$.
Now, imagine that we allow the particle to evolve as
$\lambda$ is switched infinitely slowly from 0 to 1.
For this system the quantity $H_\lambda^{3/2}{\cal V}_\lambda$
is an adiabatic invariant\cite{brown}, therefore the total
work performed on a particle with initial energy $E_0$
is: $W_\infty=[({\cal V}_0/{\cal V}_1)^{2/3} - 1]E_0$.
Taking an ensemble of such particles, defined by a 
canonical distribution of initial conditions 
(at $\lambda=0$), we get 
\begin{equation}
\lim_{t_s\rightarrow\infty}\rho(W,t_s) = 
\Biggl(
{4\beta^3 W\over\pi r^3}\Biggr)^{1/2}
\exp\Biggl(-{\beta W\over r}\Biggr) 
\theta(W),
\end{equation}
where $r=({\cal V}_0/{\cal V}_1)^{2/3}-1$.
As with the example of the harmonic oscillator, it is 
straightforward to show that this distribution $\rho$
satisfies Eq.\ref{eq:iden}.

\subsection{Langevin evolution}

Next, let us consider modifying
Hamilton's equations, by adding
both a frictional and a stochastic force:
\begin{mathletters}
\label{eq:langevin}
\begin{eqnarray}
\dot x &=& p \\
\dot p &=& -\partial_x V_\lambda
-\gamma p + \tilde F(t).
\end{eqnarray}
\end{mathletters}
We have assumed a one-degree-of-freedom system, 
and a Hamiltonian
$H_\lambda=p^2/2 + V_\lambda(x)$.
Let us furthermore take the stochastic force,
$\tilde F(t)$, to represent white noise, 
with an autocorrelation function given by
\begin{equation}
\langle\tilde F(t_1)\tilde F(t_2)\rangle
= D_P\,\delta(t_2-t_1).
\end{equation}
Finally, let us impose a fluctuation-dissipation 
relation between the frictional and stochastic forces:
\begin{equation}
\label{eq:fd}
\gamma=\beta D_P/2.
\end{equation}
An ensemble of trajectories evolving under the
stochastic process just defined
satisfies the Fokker-Planck equation,
\begin{equation}
\label{eq:fp}
{\partial f\over\partial t} =
\{H_\lambda,f\} + \gamma{\partial\over\partial p}
(pf) + {D_P\over 2}
{\partial^2 f\over\partial p^2}.
\end{equation}
Since Eq.\ref{eq:fp} is of the form
$\partial f/\partial t = \hat R_\lambda f$,
this process is Markovian.
Furthermore, inspection reveals that $\hat R_\lambda$
satisfies Eq.\ref{eq:stat}, i.e. detailed balance holds.
Thus, the conditions for the validity of Eq.\ref{eq:iden}
are met: given an ensemble of systems evolving under
this stochastic process, as $\lambda$ is changed from 0
to 1 over a time $t_s$, and given an initial distribution
$f({\bf z},0)=Z_0^{-1}\exp -\beta H_0({\bf z})$,
we are guaranteed that this ensemble will satisfy
$\overline{\exp -\beta W} = \exp -\beta\Delta F$,
for any switching time $t_s$.

Eq.\ref{eq:langevin} is a {\it Langevin} set of equations.
If $\lambda$ is held fixed, an ensemble of trajectories
evolving under these stochastic
equations of motion will relax to a canonical
distribution on phase space.
In other words, the evolution specified by
Eqs.\ref{eq:langevin} to \ref{eq:fd} satisfies not
only detailed balance, but the stronger assumption
of thermalization as well.
If we start with a canonical distribution, at
$\lambda=0$, and then switch the value of the external
parameter infinitely slowly to $\lambda=1$, then
the ensemble of trajectories will remain in quasi-static
equilibrium over the course of the switching
process:
\begin{equation}
\label{eq:quasi-stat}
f({\bf z},t) = Z_\lambda^{-1}
\exp -\beta H_\lambda({\bf z})\qquad,\qquad
\lambda=\lambda(t)\qquad,\qquad t_s\rightarrow\infty.
\end{equation}
The creeping value of $\lambda$ thus drags 
$f({\bf z},t)$ through a continuous
sequence of canonical distributions.
Furthermore, in this limit, the ``work accumulated''
will be the same function of time for every trajectory 
in the ensemble: $w(t)=F_\lambda-F_0$;
each trajectory represents a system evolving
in quasi-static equilibrium with the reservoir.
Thus, the quantity $Q({\bf z},t)$ defined earlier 
is given, in this quasi-static limit, by
\begin{equation}
\label{eq:q}
Q({\bf z},t) = 
\exp -\beta (F_\lambda-F_0) = 
Z_\lambda/Z_0.
\end{equation}
It then follows that the function $g\equiv fQ$
(Eq.\ref{eq:gdef}) satisfies
$g({\bf z},t)=Z_0^{-1}\exp -\beta H_\lambda({\bf z})$
(Eq.\ref{eq:goft}),
from which we get 
$\overline{\exp -\beta W}=\exp -\beta\Delta F$.
Of course, when we switch $\lambda$ from 0 to 1 over
a {\it finite} time $t_s$, neither Eq.\ref{eq:quasi-stat}
nor Eq.\ref{eq:q} will in general hold;
nevertheless, Eq.\ref{eq:goft}, and therefore our
central result, will remain valid, since
Eq.\ref{eq:dgdt} governs the time evolution of $g$
regardless of how slowly or quickly we switch the
external parameter.

\subsection{Isothermal molecular dynamics}
\label{subsec:imd}

The Langevin equations above provide a simple method
for numerically simulating the evolution of a 
thermostatted system
(i.e. a system in contact with a heat bath),
without explicitly simulating the many
degrees of freedom of the bath.
The term $\tilde F(t)$ may be implemented by generating
a small, random momentum ``kick'' at each time step in
the numerical integration of the equations of motion.
This method works both for static Hamiltonians and
--- as in the case considered in the present paper ---
for Hamiltonians made time-dependent through the
variation of an external parameter.
In the past decade or so, methods for using explicitly
{\it deterministic} equations of motion to simulate
thermostatted systems, have proven 
very useful\cite{hoover}.
These generally go under the name of 
{\it isothermal molecular dynamics} (IMD).
Typically, the heat bath
is represented by one or more ``extra''
degrees of freedom.
Then, in the extended phase space which includes both
the system of interest and the extra degree(s) of
freedom, the evolution is governed by a set of
deterministic (but non-Hamiltonian)
equations of motion.
These are tailored so that, when the Hamiltonian
describing the system of interest
is static, the variables representing the system
of interest explore phase space canonically,
at least to a good approximation.

An example of an IMD scheme is 
{\it Nos\' e-Hoover dynamics}\cite{n,h}, represented 
by the following equations of motion:
\begin{mathletters}
\label{eq:nh}
\begin{eqnarray}
\Bigl\{
\dot q = p\,&,&\,
\dot p = -\nabla V_\lambda-\zeta p
\Bigr\}_n\\
\label{eq:nh2}
\dot\zeta &=&
(K/K_0-1)/\tau^2.
\end{eqnarray}
\end{mathletters}
We have again assumed a kinetic-plus-potential 
Hamiltonian;
the index $n$ runs over all $D$ degrees of
freedom of the system, $K=\sum_n p_n^2/2$ is 
the total kinetic energy of the system,
$K_0=D/2\beta$ is the thermal average of $K$, and
the parameter $\tau$ acts as a relaxation time.
Let us imagine a trajectory evolving in the
extended phase space, 
$({\bf z},\zeta)$-space, under these equations
of motion, as $\lambda$ is switched from 0 to 1.
The work $W$ performed on the system of interest
is defined, as before, by Eq.\ref{eq:work}.
While this expression does not explicitly contain the
bath variable $\zeta$, $W$ is nevertheless a function
of the full set of initial conditions,
$({\bf z}_0,\zeta_0)$, since the evolution of the
system of interest ({\bf z}) is coupled to that of
the heat bath ($\zeta$).

In Ref.\cite{iden}, it was shown that, if one
started with a distribution of initial conditions
in the extended phase space given by
\begin{equation}
f({\bf z},\zeta,0) \propto
\exp -[\beta H_0({\bf z})+
\zeta^2\tau^2/2],
\end{equation}
and if one then propagated an ensemble of trajectories
from these initial conditions, under the
Nos\' e-Hoover (NH) equations of motion, as $\lambda$
was switched from 0 to 1 over a time $t_s$, and
computed the work $W$ for each trajectory, then
Eq.\ref{eq:iden}
would hold identically for this ensemble.
This was shown by inspection.

Our purpose here is to use the central result of the
present paper to establish simple criteria for 
determining whether or not Eq.\ref{eq:iden} is valid
for a particular implementation of IMD.
In general, the heat bath is represented by $N$ 
variables, $\zeta_1$ to $\zeta_N$.
Let ${\bf y}=({\bf z},\zeta_1,\cdots,\zeta_N$) denote
a point in the $(2D+N)$-dimensional extended phase
space, where $D$ is the number of degrees of freedom
of the system of interest.
Then the thermostatting scheme in question is defined
by a set of deterministic equations of motion in this
space:
\begin{equation}
\label{eq:scheme}
\dot{\bf y} = {\bf K}_{\lambda}({\bf y}),
\end{equation}
of which Eq.\ref{eq:nh} is an example.
An ensemble of trajectories evolving under
these equations of motion may be
described by a distribution $f({\bf y},t)$ which
satisfies the continuity equation,
\begin{equation}
\label{eq:nhpde}
{\partial f\over\partial t} = -{\partial\over\partial{\bf y}}
\cdot({\bf K}_\lambda f)
\equiv \hat R_\lambda f.
\end{equation}

Now suppose that we can find a function 
$q(\zeta_1,\cdots,\zeta_N)$ such that the distribution
\begin{equation}
f_\lambda^C\propto\exp -\beta[
H_\lambda({\bf z})+q(\zeta_1,\cdots,\zeta_N)]
\end{equation}
is stationary under the evolution defined by 
Eq.\ref{eq:scheme}, when $\lambda$ 
is held fixed.
That is, 
\begin{equation}
\label{eq:nhdb}
\hat R_\lambda f_\lambda^C = 0.
\end{equation}
(For the NH equations, $q=\zeta^2\tau^2/2\beta$ 
satisfies this condition.)
We will allow $q$ to depend on $\beta$, and any other
{\it constant} parameters, but not on $\lambda$.
The distribution $f_\lambda^C({\bf y})$ may be viewed as
the ``canonical'' distribution in the extended phase
space, since it is invariant when $\lambda$ is held
fixed.
Now define a function
\begin{equation}
{\cal H}_\lambda({\bf y}) \equiv
H_\lambda({\bf z}) + q(\zeta_1,\cdots,\zeta_N),
\end{equation}
which we may think of as an {\it extended Hamiltonian}.
(This is not meant to imply that ${\cal H}_\lambda$
generates Eq.\ref{eq:scheme} as a ``real'' Hamiltonian
would; in general those equations are non-Hamiltonian.)
By Eq.\ref{eq:nhdb}:
\begin{equation}
\label{eq:nhdb2}
\hat R_\lambda\exp -\beta{\cal H}_\lambda({\bf y})=0.
\end{equation}

Now forget for a moment the division between system of
interest and heat bath, and treat the entire
extended phase space as a phase space for some system
of interest, for which a parameter-dependent energy function,
${\cal H}_\lambda({\bf y})$, is defined.
The evolution given by Eq.\ref{eq:scheme} then 
satisfies the two conditions listed in 
Section \ref{sec:deriv}:
(1) it is Markovian (Eq.\ref{eq:nhpde}), and
(2) it satisfies detailed balance 
(Eq.\ref{eq:nhdb2}).
Thus, our central result, 
$\overline{\exp -\beta W}=\exp -\beta\Delta F$,
is identically true for an ensemble of trajectories
evolving under Eq.\ref{eq:scheme} from an initial distribution 
$f_0\propto\exp -\beta{\cal H}_0$,
{\it provided we replace $H_\lambda({\bf z})$ by 
${\cal H}_\lambda({\bf y})$ in computing
$W$ and $\Delta F$}.
However, it is easily verified that
we will get exactly the same
values for both $W$ and $\Delta F$
using ${\cal H}_\lambda({\bf y})$, as we would 
obtain with $H_\lambda({\bf z})$.
In the case of $W$, this is because
only the first term of ${\cal H}_\lambda$ 
(namely, $H_\lambda$) depends on $\lambda$;
for $\Delta F$, it follows from the fact that the
partition function for the extended Hamiltonian
factorizes into an integral over the ${\bf z}$ 
variables and an integral over the $\zeta$ variables,
and only the former depends on $\lambda$.
(See the definitions of work and free energy, 
Eqs.\ref{eq:fdef} and \ref{eq:work}.)
These considerations lead to the following 
simple conclusion.

Suppose we have a system described by a 
Hamiltonian $H_\lambda({\bf z})$, and we wish to 
compute the free energy difference $\Delta F=F_1-F_0$.
Suppose furthermore that we are given a scheme for
IMD; that is, we have a set of ``heat
bath'' variables $(\zeta_1,\cdots,\zeta_N)$, along with
equations of motion in the extended phase space,
as per Eq.\ref{eq:scheme}.
Then, if a function $q$ of the heat bath variables
can be found such that the distribution 
$f_\lambda^C\propto\exp -\beta(H_\lambda+q)$ is stationary
under the equations of motion (with $\lambda$ fixed),
then we can compute $\Delta F$ as follows.  
Let an ensemble of trajectories evolve from an 
initial distribution $f_0\propto\exp -\beta(H_0+q)$,
as $\lambda$ is switched from 0 to 1 over a finite
switching time $t_s$.
Compute the work $W$ for each trajectory, as per
Eq.\ref{eq:work}, and then compute the ensemble
average of $\exp -\beta W$.
This average will equal $\exp -\beta\Delta F$.

\subsection{Monte Carlo evolution}

As a final example --- again, motivated by 
computer simulations --- let us consider the evolution
of a {\it Monte Carlo} (MC) trajectory,
as $\lambda$ is switched from 0 to 1.
In this case both the trajectory and the 
parameter $\lambda$ evolved in discrete steps, rather
than continuously:
\begin{eqnarray}
{\bf z}(t) &\rightarrow& {\bf z}_0,{\bf z}_1,\cdots,
{\bf z}_N \\
\lambda(t) &\rightarrow& \lambda_0,\lambda_1,\cdots,
\lambda_N \qquad ;\qquad \lambda_n=n/N.
\end{eqnarray}
The initial point in phase space, ${\bf z}_0$, is
sampled from a canonical distribution, with the
value of $\lambda$ fixed at $\lambda_0=0$.
One then imagines that $\lambda$ changes 
abruptly from $\lambda_0=0$ to $\lambda_1=1/N$;
as a result, a quantity of work 
$\delta W_1 = H_{\lambda_1}({\bf z}_0) - 
H_{\lambda_0}({\bf z}_0)$ is performed on the system.
Then, the system jumps to the next point in 
phase space, ${\bf z}_1$, generated from 
${\bf z}_0$ by a MC algorithm appropriate to 
the Hamiltonian $H_{\lambda_1}$
(see Appendix B).
One then continues to alternate between 
discrete changes in $\lambda$ and discrete MC
jumps in phase space, until the entire
``trajectory'' (${\bf z}_0,\cdots,{\bf z}_N$) 
is obtained, and the value of $\lambda$ is 1.
The total work performed during this discrete
switching process is:
\begin{equation}
\label{eq:dwork}
W = \sum_{n=1}^N \delta W_n =
\sum_{n=1}^N \Bigl[ H_{\lambda_n}({\bf z}_{n-1}) -
H_{\lambda_{n-1}}({\bf z}_{n-1})\Bigr].
\end{equation}
Note that ``time'' does not
enter into this scheme.
The quasi-static limit is 
obtained by letting the number of steps
become arbitrarily large: $N\rightarrow\infty$.

Let us now imagine that we generate an infinite
ensemble of MC trajectories, each of length $N$.
We do this by implementing the above procedure
repeatedly, each time feeding in a different 
string of random numbers to generate the initial
conditions ${\bf z}_0$, and the subsequent MC
steps.
Given such an ensemble, we compute the work $W$
performed on each trajectory (Eq.\ref{eq:dwork}),
and then the ensemble average of $\exp -\beta W$.
As shown in Appendix B, this average will equal
$\exp -\beta\Delta F$.
This should come as no surprise:
a trajectory generated by the MC algorithm is
a {\it Markov chain}, with detailed balance
built into the individual steps.
This evolution thus satisfies, in discretized form,
the two assumptions required for the validity of
Eq.\ref{eq:iden}.

Let us now consider the two limiting cases
$N=1$ and $N\rightarrow\infty$.
The latter is the MC equivalent of the quasi-static
limit.
In this case our ensemble proceeds through a discrete,
infinitesimally spaced sequence of canonical 
equilibrium distributions, and the work performed
on each trajectory is $\Delta F$, as per Eq.\ref{eq:infty}.
The result $\overline{\exp -\beta W}=\exp -\beta\Delta F$
then follows automatically.
In the opposite limit, $N=1$, the work $W$ performed
on a particular ``trajectory'' is given by
\begin{equation}
W = H_1({\bf z}_0) - H_0({\bf z}_0) \equiv
\Delta H({\bf z}_0) \qquad (N=1).
\end{equation}
Eq.\ref{eq:iden} then reduces to:
\begin{equation}
\langle\exp -\beta\Delta H\rangle_0 
= \exp -\beta\Delta F,
\end{equation}
where $\langle\cdots\rangle_0$ denotes a canonical
average with respect to $\lambda=0$.
This result is a well-known identity for 
the free energy difference $\Delta F$;
see Eq.\ref{eq:tp} below.

\section{FREE ENERGY COMPUTATIONS}
\label{sec:comp}

While Eq.\ref{eq:iden} is interesting in its own
right, it may additionally prove useful in the
numerical computation of free energy differences.
The field of free energy computations is decades
old, with diverse applications, and a very large 
body of literature exists on the subject\cite{reviews}.
In this section, without attempting a survey
of the field, we discuss a few points relevant
to the possible application of Eq.\ref{eq:iden}
to free energy computations.
These comments expand on ones made in Ref.\cite{iden}.

Most methods of computing free energy differences
are variants of either the {\it thermodynamic
integration} (TI) or {\it thermodynamic perturbation}
(TP) methods.
(For an exception to this statement, see the
work of Holian, Posch, and Hoover\cite{hph},
where two new expressions for $\Delta F$, based
on time-integrated heat transfer, are derived
within the framework of isothermal molecular
dynamics.)
TP is based on the identity\cite{tp}
\begin{equation}
\label{eq:tp}
\Delta F = -\beta^{-1}\ln
\langle\exp -\beta\Delta H\rangle_0
\qquad,\qquad (\Delta H\equiv H_1-H_0),
\end{equation}
where $\langle\cdots\rangle_\lambda$ denotes
a canonical average with respect to a fixed 
value of $\lambda$.
Using a method such as Monte Carlo, one samples
$N$ points in phase space from the canonical 
distribution corresponding to $\lambda=0$,
and then one takes the average of 
$\exp -\beta\Delta H$ over these $N$
points.
In principle, the method is exact for 
$N\rightarrow\infty$;
in practice, unless the canonical distributions
corresponding to $H_0$ and $H_1$ overlap to
a significant degree,
the average of $\exp -\beta\Delta H$
will be dominated by points in phase space 
which are visited extremely rarely during the
canonical sampling, so numerical convergence
with $N$ will be prohibitively slow.
One way to get around this problem is to
break up the $\lambda$-interval $[0,1]$
into small sub-intervals,
then use Eq.\ref{eq:tp} to compute the free
energy difference corresponding to each 
sub-interval.
Other, more sophisticated methods of extracting
the best efficiency from Eq.\ref{eq:tp} have
been developed over the years\cite{bennett}.

TI is based on the identity\cite{ti} 
\begin{equation}
\label{eq:ti}
\Delta F = \int_0^1 d\lambda\,
\Bigl\langle{\partial H_\lambda\over\partial\lambda}
\Bigr\rangle_\lambda.
\end{equation}
The integral on the right may be evaluated by
sampling $n_s$ points in phase space from each
of $M$ different canonical distributions
(corresponding to equally spaced values of
$\lambda$ from 0 to 1), for a grand total of
$N=n_sM$ points sampled.
The average of $\partial H_\lambda/\partial\lambda$
is then computed at each value of $\lambda$, 
and from these $M$ averages the integral is obtained.
With the exception of a few very simple systems
(e.g.\ ideal gases, harmonic oscillators),
the standard way to obtain a canonical distribution
is to first allow the system to ``age'' ---
that is, to relax to an equilibrium statistical
state --- under some MC or IMD scheme.
In implementing the numerical evaluation of
$\int_0^1 d\lambda\,\langle
\partial H_\lambda/\partial\lambda\rangle_\lambda$,
it is often too
time-consuming to age the system independently 
at each of the $M$ selected values of $\lambda$.
Instead, the final point sampled at one $\lambda$
value may be used to generate the initial point 
sampled at the next value of $\lambda$.
Of course, this means that at each $\lambda$
(except $\lambda=0$) we are actually sampling
from a slightly out-of-equilibrium distribution,
which leads to a systematic error (in fact, an
over-estimate) in the evaluation of the integral.

A limiting case of this procedure arises when
we take $n_s=1$, i.e.\ exactly one point is
sampled at each value of $\lambda$.
This is the {\it slow growth} method\cite{sg}.
As stressed by Reinhardt and 
coworkers\cite{wprjeh,wata,hrd},
if we view the chain of points 
$({\bf z}_0,\cdots,{\bf z}_N)$ thus generated
as a {\it trajectory}
evolving in phase space (as $\lambda$ is switched
from 0 to 1), then 
the integral appearing on the right side of
Eq.\ref{eq:ti} represents the work $W$
performed on the system over the course of
the switching process.
This picture is a compelling one, as it attaches
a very physical interpretation to the numerical
evaluation of Eq.\ref{eq:ti}:
instead of computing an integral, we are
simulating the evolution of a thermostatted
system, with the idea that, in the limit
of infinitely slow switching,
the work performed on the system will equal the 
free energy difference $\Delta F$.
For a finite switching rate, the above-mentioned
systematic error inherent in the slow
growth method is simply a manifestation of
the inequality $\overline{W}\ge\Delta F$;
see Eq.\ref{eq:ineq} and Refs.\cite{wprjeh,hrd,taiwan}.
This interpretation of free energy computations 
is referred to as {\it adiabatic switching}
(or {\it finite-time variation}), and indeed 
may be viewed as a separate method, distinct
from thermodynamic integration.

In the context of adiabatic switching, Eq.\ref{eq:iden}
says that, if we run an ensemble of
finite-time (or finite-$N$, for Monte Carlo) simulations of 
the switching process, using for instance one of the methods
described in Section \ref{sec:examp}, then the
average of $\exp -\beta W$ over this ensemble of
simulations will equal $\exp -\beta\Delta F$.
Assuming perfect numerical accuracy and an infinite 
ensemble of simulations, this is an {\it exact} statement.
Another way of putting it is as follows:
for a single finite-time switching simulation,
the value of $\exp -\beta W$ provides
an {\it unbiased} estimate of $\exp -\beta\Delta F$.
By contrast,
the value of $W$ gives a biased estimate
of $\Delta F$ (i.e.\ $\overline{W}\ge\Delta F$).
Indeed, the former statement implies
the latter, in the same way that 
Eq.\ref{eq:tp} implies the 
Gibbs-Bogoliubov-Feynman inequality\cite{chandler}:
$\langle\Delta H\rangle_0\ge\Delta F$.
Let us consider this for a moment.

Given a real function of a real variable,
$y(x)$, it is easy to show that, if 
$d^2y/dx^2\ge 0$ for all $x$, then
\begin{equation}
\label{eq:math}
{1\over N}\sum_{i=1}^N y(x_i) \ge
y\Bigl( {1\over N}\sum_{i=1}^N x_i\Bigr),
\end{equation}
for any set of points $\{x_1,\cdots,x_N\}$.
If these points are the result of random
sampling from some ensemble, then Eq.\ref{eq:math},
in the limit $N\rightarrow\infty$, may be
rewritten as
\begin{equation}
\overline{y(x)} \ge y(\overline{x}),
\end{equation}
where the overbar denotes an ensemble average.
Applying this result to $y(x) = \exp x$,
with $x=-\beta W$, we get
\begin{equation}
\overline{\exp -\beta W} \ge 
\exp -\beta\overline{W}.
\end{equation}
This is identically true for any distribution
$\rho(W)$.
We may now combine this result with Eq.\ref{eq:iden}
to get $\overline{W}\ge\Delta F$.

Now, what if we perform a finite number, $N_s$,
of identical switching simulations?
Let $W_i$ denote the work performed on the
system during the $i$'th simulation, and let
\begin{equation}
\label{eq:wadef}
W^a\equiv{1\over N_s}\sum_{i=1}^{N_s} W_i
\end{equation}
be the average over these values.
We may view the $W_i$'s as numbers sampled
randomly from a distribution $\rho(W)$ 
satisfying Eq.\ref{eq:iden}.
Then the expectation value of $W^a$ provides
a rigorous upper bound on $\Delta F$:
\begin{equation}
\langle\langle W^a \rangle\rangle =
\overline{W} \equiv
\int dW\,\rho(W)\,W \ge\Delta F.
\end{equation}
The double angular brackets, denoting
{\it expectation value}, specifically mean an 
average over all possible sets of $N_s$
simulations.
Now, Eq.\ref{eq:iden} suggests that, rather
than $W^a$, we consider the following quantity
as our best estimate of $\Delta F$:
\begin{equation}
\label{eq:wxdef}
W^x \equiv -\beta^{-1}\ln
{1\over N_s}\,\sum_{i=1}^{N_s}
\exp -\beta W_i.
\end{equation}
For $N_s=1$, $W^x$ and $W^a$ are identical,
and the expectation value of either is 
$\overline{W}$.
For $N_s\rightarrow\infty$, by contrast, 
$W^x$ converges to $\Delta F$,
whereas $W^a$ converges to $\overline{W}$.
For intermediate values of $N_s$, the following
inequality chain holds:
\begin{equation}
\label{eq:chain}
\Delta F\le
\langle\langle W^x \rangle\rangle \le
\langle\langle W^a \rangle\rangle.
\end{equation}
(Both inequalities are derived by combining
Eqs.\ref{eq:iden} and \ref{eq:math} with 
the definitions of $W^a$ and $W^x$.)
In other words, as an estimate of $\Delta F$,
the ``exponential average'', $W^x$, is
statistically less biased than the ordinary
average, $W^a$, for $N_s>1$.

On the face of it, the last statement seems to imply
that, if we perform more than one switching
simulation, then we are better off using $W^x$
rather than $W^a$ as our best guess (or upper
bound) for $\Delta F$.
In practice, however, Eq.\ref{eq:iden}
may be subject to the same disease as the TP
identity, Eq.\ref{eq:tp}.
Namely, if the values of $W$ obtained from 
repetitions of the switching simulation
typically differ from one another by much
more than $\beta^{-1}$, then the average of
$\exp -\beta W$ will be dominated by values
of $W$ which are very rarely sampled\cite{disease}.
Thus, the convergence of $W^x$ to $\Delta F$,
in the limit $N_s\rightarrow\infty$, may be
much slower than the convergence
of $W^a$ to $\overline{W}$, in the same
limit.
In other words, for a finite number of switching
simulations, $W^x$ may be subject to considerably 
larger {\it statistical} fluctuations than $W^a$, 
even though its {\it systematic} error (expectation 
value minus $\Delta F$) is, by Eq.\ref{eq:chain}, 
smaller.

The preceding comments point to the following
tentative conclusion.
If one runs a set of $N_s$ switching simulations,
with the goal of computing $\Delta F$,
and if the spread in the values of work $W$
obtained is not much larger than $\beta^{-1}$,
then the exponential average $W^x$ defined by
Eq.\ref{eq:wxdef} should provide a better estimate of
(or tighter upper bound on) $\Delta F$ than the
ordinary average $W^a$.
This conclusion is supported by a calculation
by John E.\ Hunter III, as described in Ref.\cite{iden};
see also the numerical illustrations in the 
following section.

Of course, a more detailed study of the possible
utility of Eq.\ref{eq:iden} to free energy 
computations should be made.
In particular, it is not ruled out that 
there exist methods around the limitation
mentioned in the previous paragraphs\cite{frenkel}.

\section{NUMERICAL RESULTS}
\label{sec:numres}

In this section we illustrate our central result
with numerical experiments.
The first four sets of simulations involve a
harmonic oscillator whose natural frequency is
switched from $\omega_0=1.0$ to $\omega_1=2.0$.
The evolution is implemented using, in turn,
each of the four examples discussed in Section
\ref{sec:examp}. 
Then, we present results involving a more complicated
system: a gas of interacting particles inside an
externally pumped piston.
All of these cases satisfy the condition discussed
at the end of Section \ref{sec:comp},
namely, the spread in the values of $W$
is not much greater than $\beta^{-1}$.
(Otherwise, the convergence of $W^x$ to $\Delta F$, 
in the limit of many simulations, would be poor.)

For the harmonic oscillator simulations, we 
take the Hamiltonian given by Eq.\ref{eq:hoham},
with $\omega_\lambda = 1.0 + \lambda$,
and $\lambda(t)=t/t_s$ as throughout this paper.
Also, we take $\beta^{-1} = 1.5$. 
Thus, the free energy difference is
$\Delta F=\beta^{-1}\ln(\omega_1/\omega_0)=1.0397$.

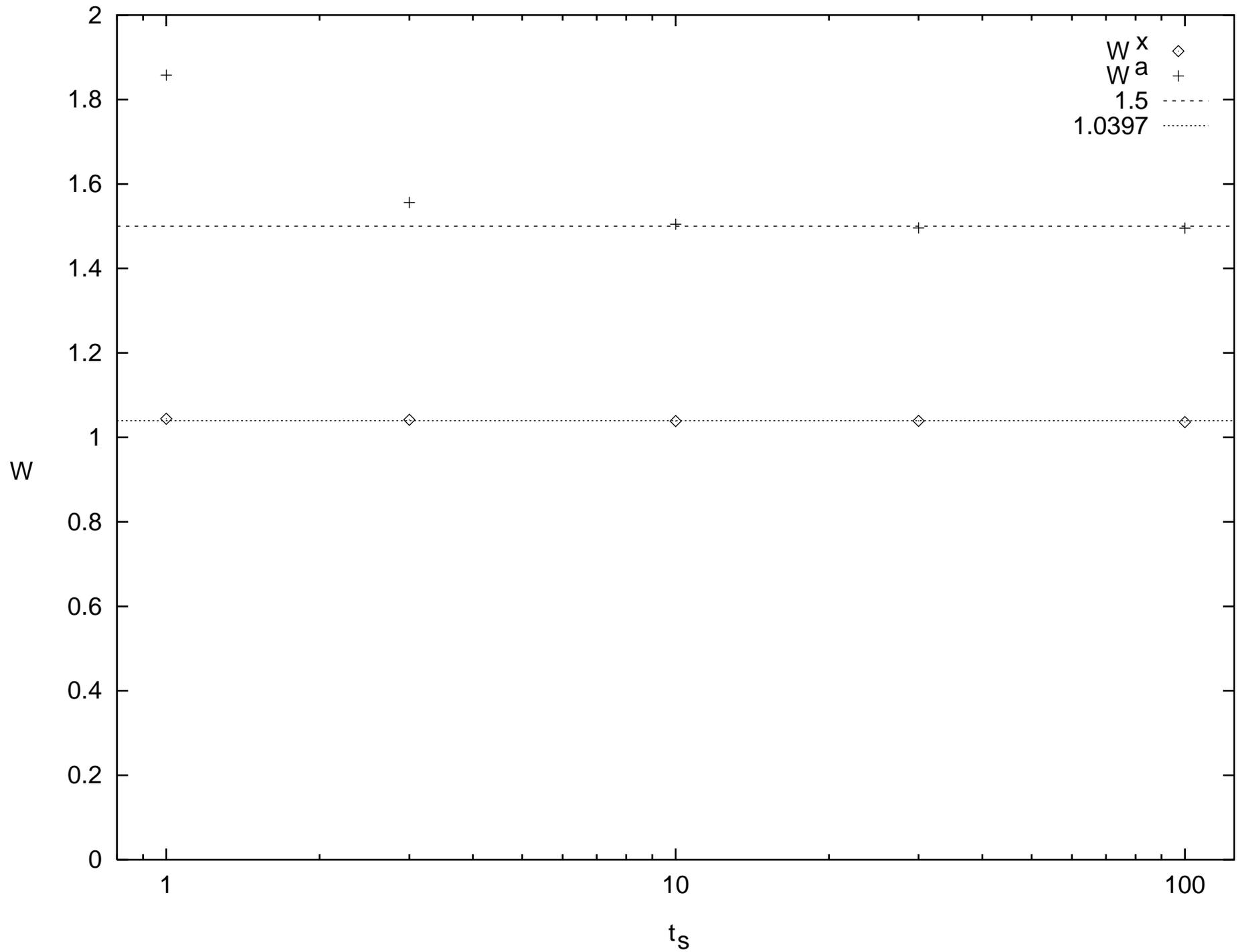
\begin{figure}
\caption{
Simulations of an isolated harmonic oscillator
whose natural frequency is switched from 
$\omega_0=1.0$ to $\omega_1=2.0$ over a switching
time $t_s$.
At each of five values of $t_s$, $10^5$ simulations
were carried out.
The upper and lower sets of points show the 
ordinary averages ($W^a$) and the exponential
averages ($W^x$) of the work, respectively.
The dashed line is at $W=1.5$, the dotted line
at $W=\Delta F=1.0397$.}
\label{fig:ham}
\end{figure}

In the first set of simulations, the oscillator
is isolated (it evolves under Hamilton's equations,
with $\lambda$ time-dependent), although the
initial conditions are sampled from a canonical
ensemble corresponding to $\lambda = 0$:
\begin{equation}
f(x,p,0) = {\beta\omega_0\over 2\pi}
\exp [-\beta(p^2+\omega_0^2 x^2)/2].
\end{equation}
Five different values of the switching time
were chosen: $t_s = $ 1.0, 3.0, 10.0, 30.0, and 100.0,
and for each $t_s$ a total of $N_s=10^5$ 
simulations were carried out.
Fig.\ref{fig:ham} shows the average value of work 
obtained at each switching time, $W^a$, as well as 
the ``exponential average'', $W^x$.
(See Eqs.\ref{eq:wadef} and \ref{eq:wxdef}.)
Since Hamiltonian evolution satisfies detailed
balance, but not thermalization (as defined
in Section \ref{sec:prelim}), we do not expect
the work performed in the limit $t_s\rightarrow\infty$
to equal the free energy difference $\Delta F$.
Rather, we expect 
$W_\infty = [(\omega_1/\omega_0) - 1] E_0$ for
a single trajectory (see Section \ref{subsec:hamev}),
and therefore, for a canonical distribution of
initial energies,
\begin{equation}
\lim_{t_s\rightarrow\infty} \overline{W} = 
[(\omega_1/\omega_0) - 1] \beta^{-1} = 1.5.
\end{equation}
The values of $W^a$ shown in Fig.\ref{fig:ham} are
consistent with this expectation.
Eq.\ref{eq:iden}, meanwhile, predicts
\begin{equation}
\lim_{N_s\rightarrow\infty} W^x = \Delta F
\end{equation}
for {\it any} value of $t_s$.
Again, the numerical results are consistent with
the prediction:
the values of $W^x$ shown in Fig.\ref{fig:ham} 
all fall very close to $\Delta F$.

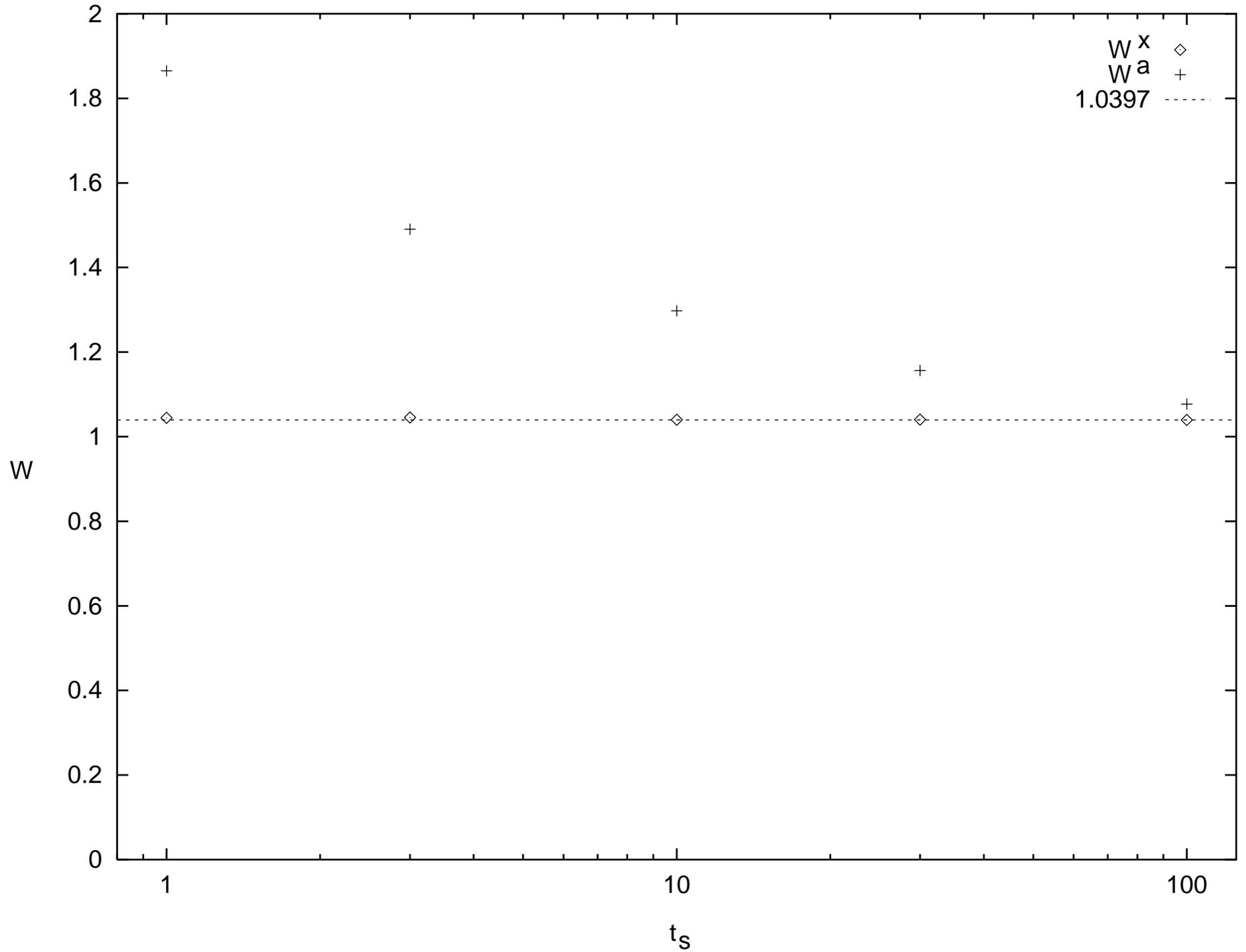
\begin{figure}
\caption{
Same as Fig.\ref{fig:ham}, except that the harmonic
oscillator is now subject to a frictional and
a stochastic force, as per Eq.\ref{eq:langevin}.
The dashed line gives the free energy difference,
$\Delta F = 1.0397$.}
\label{fig:lang}
\end{figure}

For the second set of simulations, we added a 
frictional and a stochastic force, as
described by Eqs.\ref{eq:langevin} to \ref{eq:fd},
with $D_P = 0.6$ and $\beta^{-1} = 1.5$.
The evolution now represents that of an
oscillator coupled to a heat bath.
The stochastic force was implemented by
generating a random momentum kick (sampled from
a Gaussian distribution) at each time step,
$dt = 0.01$, in the numerical integration.
As with the Hamiltonian evolution, $10^5$ 
simulations were performed at each of the five 
switching times, and the results for 
$W^a$ and $W^x$ are plotted in Fig.\ref{fig:lang}.
Here we {\it do} expect that the work $W$
will approach $\Delta F = 1.0397$ as 
$t_s\rightarrow\infty$, and the results for $W^a$ 
support this.
At the same time, the exponential average $W^x$
falls very close to $\Delta F$ for each switching 
time, as predicted by Eq.\ref{eq:iden}.

The next simulations again involved a thermostatted
harmonic oscillator, only this time isothermal
molecular dynamics was used to implement the
coupling to the heat bath.
The particular IMD scheme used was 
developed by Hoover and Holian\cite{hh},
and involves two heat bath variables,
$\zeta$ and $\xi$.
The equations of motion in the extended phase
space are 
\begin{mathletters}
\label{eq:imdeq}
\begin{eqnarray}
\dot x &=& p \\
\dot p &=& -\omega_\lambda^2 x - \zeta p
- \beta\xi p^3 \\
\dot\zeta &=& (\beta p^2-1)/\tau^2 \\
\dot\xi &=& (\beta^2 p^4 - 3\beta p^2)/\tau^2, 
\end{eqnarray}
\end{mathletters}
where $\tau$ is a relaxation time whose
value was set to unity.
A total of $N_s=10^5$ switching simulations were
performed, with a swtiching time $t_s=1.0$.
At the start of each simulation, initial conditions
were sampled from the distribution
\begin{equation}
f(x,p,\zeta,\xi,0) \propto
\exp [-\beta(p^2+\omega_0^2 x^2)/2\,\,
-\tau^2(\zeta^2+\xi^2)/2].
\end{equation}
It is easily verified by inspection that, if
the value of $\lambda$ were held fixed at 0,
then this distribution would be invariant under
Eq.\ref{eq:imdeq}.
This is therefore the ``canonical distribution''
corresponding to this IMD scheme, and the function
$q$ defined in Section \ref{subsec:imd} is given by:
\begin{equation}
q(\zeta,\xi) = \tau^2(\zeta^2+\xi^2)/2\beta.
\end{equation} 

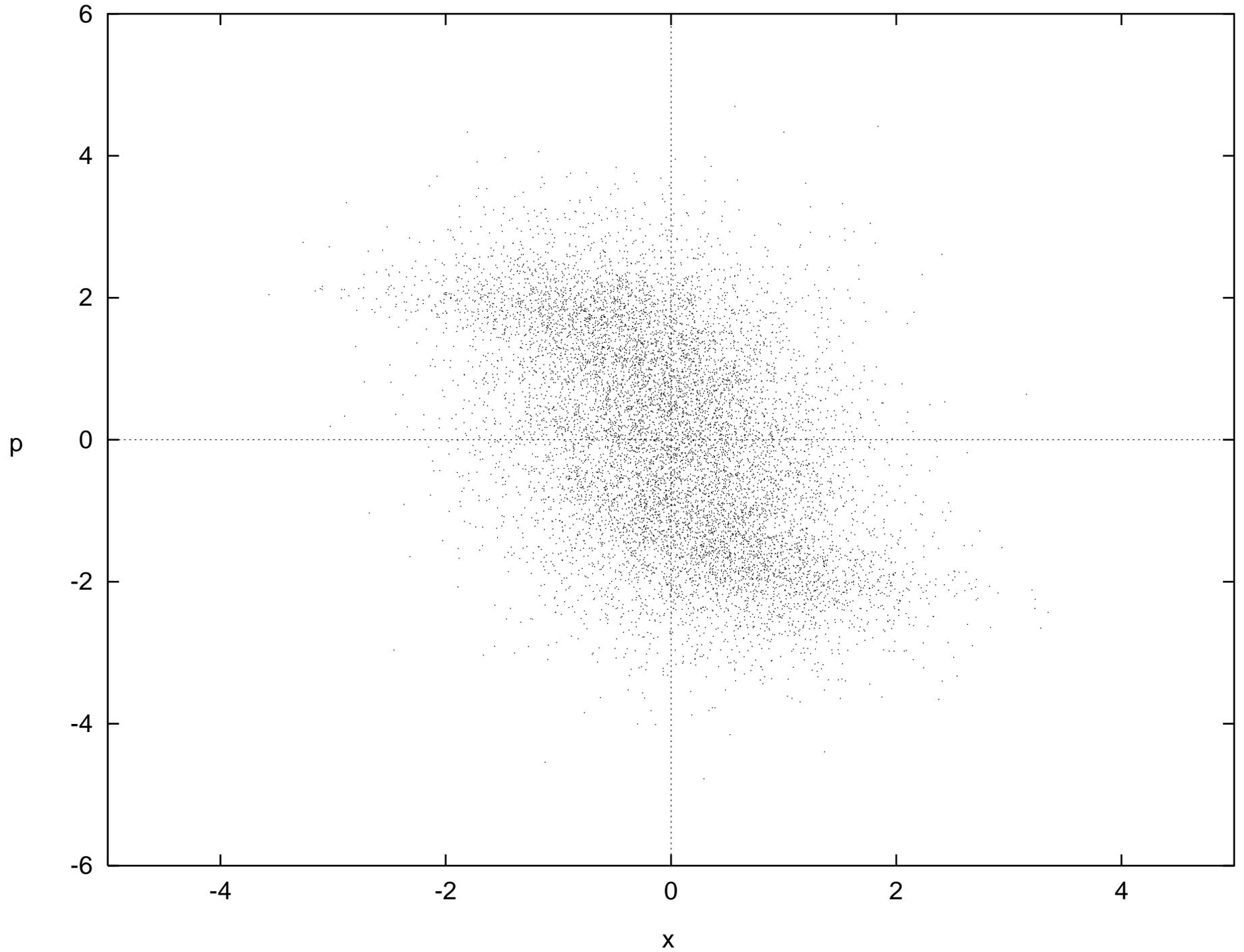
\begin{figure}
\caption{
In these simulations, the harmonic oscillator
is thermostatted with the IMD scheme described
in the text.
$10^5$ simulations at $t_s=1.0$ were performed,
and the dots in this figure show the final 
locations in phase space of these trajectories.}
\label{fig:imd1}
\end{figure}

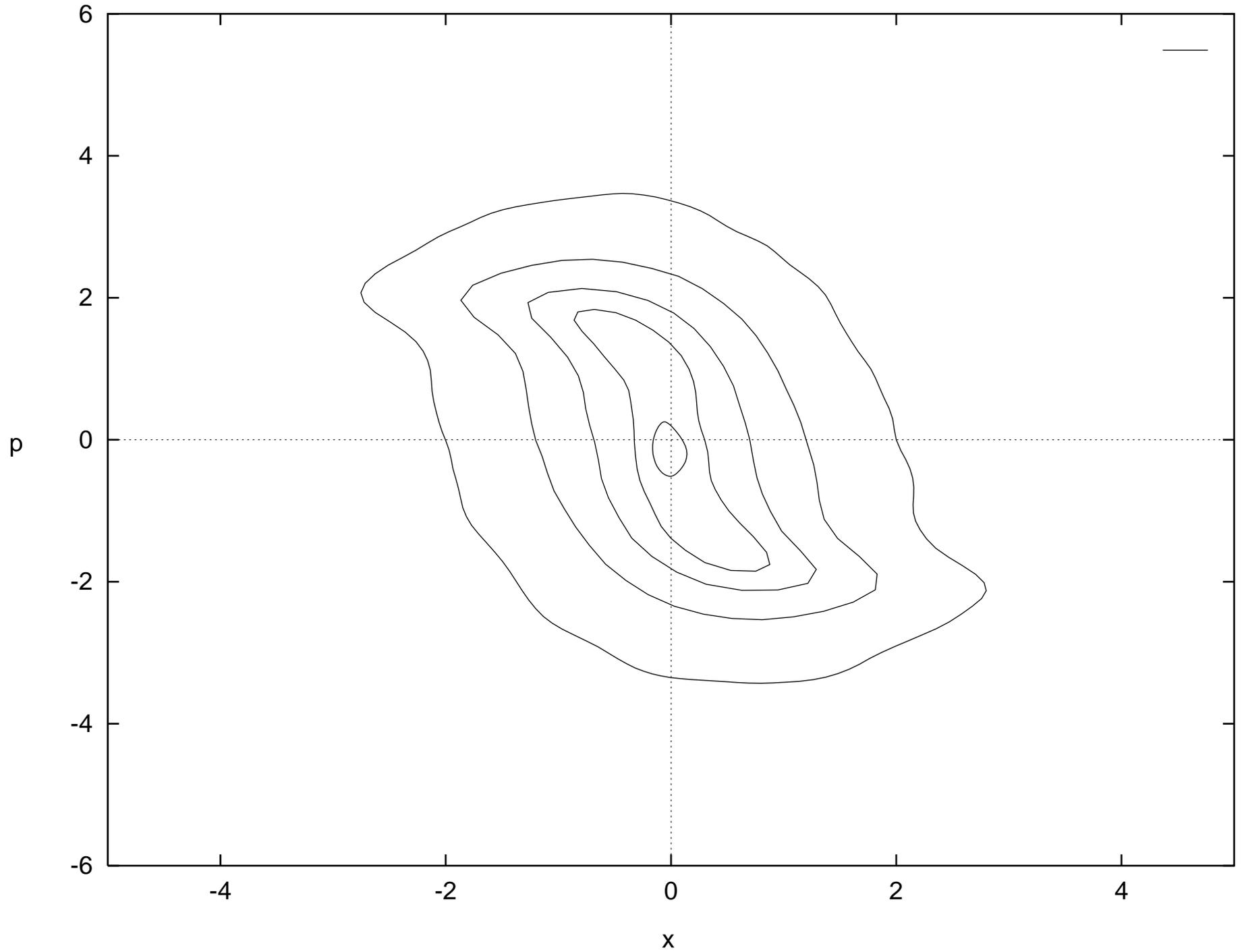
\begin{figure}
\caption{
Contour plot of the distribution $f(x,p,t_s)$,
constructed from the data shown in 
Fig.\ref{fig:imd1}, with Gaussian smoothing.}
\label{fig:imd2}
\end{figure}

This set of simulations was used to illustrate
Eq.\ref{eq:goft}, evaluated at $t=t_s$.
(See the discussion at the end of Section \ref{sec:deriv}.)
Fig.\ref{fig:imd1} is a scatter plot showing
the distribution of final values in phase space,
$(x(t_s),p(t_s))$, for the $10^5$ trajectories.
Fig.\ref{fig:imd2} shows several contour lines
of this distribution, after smearing each point
with a Gaussian of variance 0.04 in both the
$x$ and $p$ directions.
Thus, the lines shown are actually contours of 
the function
\begin{equation}
\tilde f(x,p) = 
{1\over N_s}\sum_{i=1}^{N_s}
\delta_\epsilon[x-x_i(t_s)]\,
\delta_\epsilon[p-p_i(t_s)]\qquad,\qquad 
\epsilon = 0.04,
\end{equation}
where $\delta_\epsilon$ is a normalized Gaussian
of variance $\epsilon$, and $(x_i(t),p_i(t))$
gives the phase space evolution of the $i$'th 
trajectory in the ensemble of simulations.
As can be seen from both figures, this distribution 
does not correspond to the canonical distribution 
for $H_1$.
Indeed, its skewness illustrates the ``lag'' which
develops between the evolving phase space 
density and the instantaneous canonical distribution.

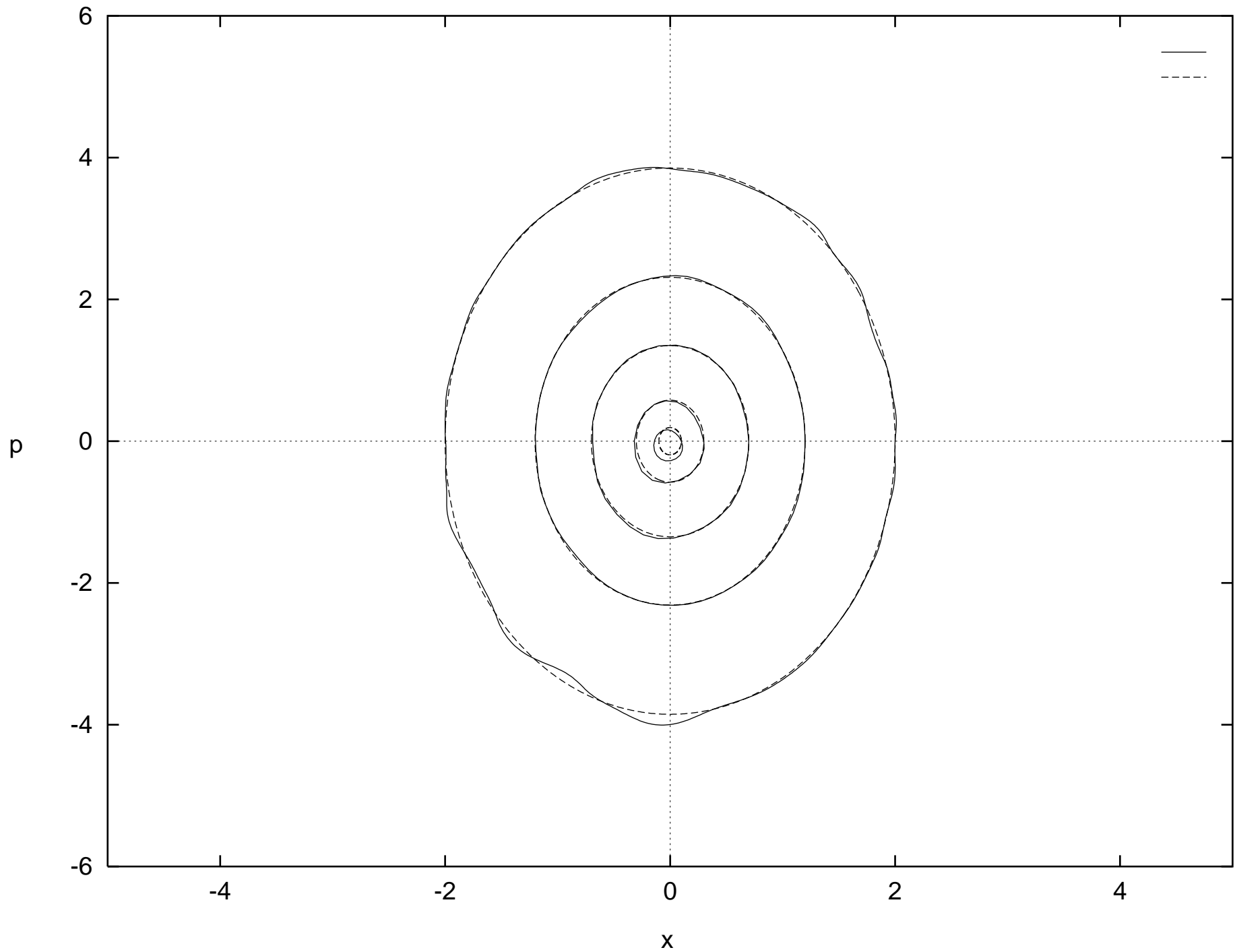
\begin{figure}
\caption{
The solid line shows contours of the 
function $g(x,p,t_s)$, constructed from
the data shown in Fig.\ref{fig:imd1};
the dashed line shows contours of the theoretical
prediction for $g(x,p,t_s)$ (Eq.\ref{eq:goft}).
Both are smoothed with Gaussians.}
\label{fig:imd3}
\end{figure}

Next, the solid lines in Fig.\ref{fig:imd3} show
contours of the function $g(x,p,t_s)$ defined in
Section \ref{sec:deriv}, obtained from the same
set of simulations.
Again, Gaussian smoothing was used, so the 
solid lines are contours of the function
\begin{equation}
\tilde g(x,p,t_s) = 
{1\over N_s}\sum_{i=1}^{N_s}
\delta_\epsilon[x-x_i(t_s)]\,
\delta_\epsilon[p-p_i(t_s)]\exp -\beta W_i
\qquad,\qquad
\epsilon = 0.04,
\end{equation}
where $W_i$ is the total work performed on the
system during the $i$'th simulation.
The dashed lines in Fig.\ref{fig:imd3} show
the corresponding contours of the {\it predicted}
``mass density'' $g(x,p,t_s)$ (Eq.\ref{eq:goft}),
with the same Gaussian smoothing function
folded in.
The agreement between the two sets of 
contours is very good.
This shows that, indeed, when one assigns a
weight $\exp -\beta W_i$ to each of the points
in the scatter plot, Fig.\ref{fig:imd1}, then 
the resulting weighted distribution is
canonical, in the sense of Eq.\ref{eq:goft}.

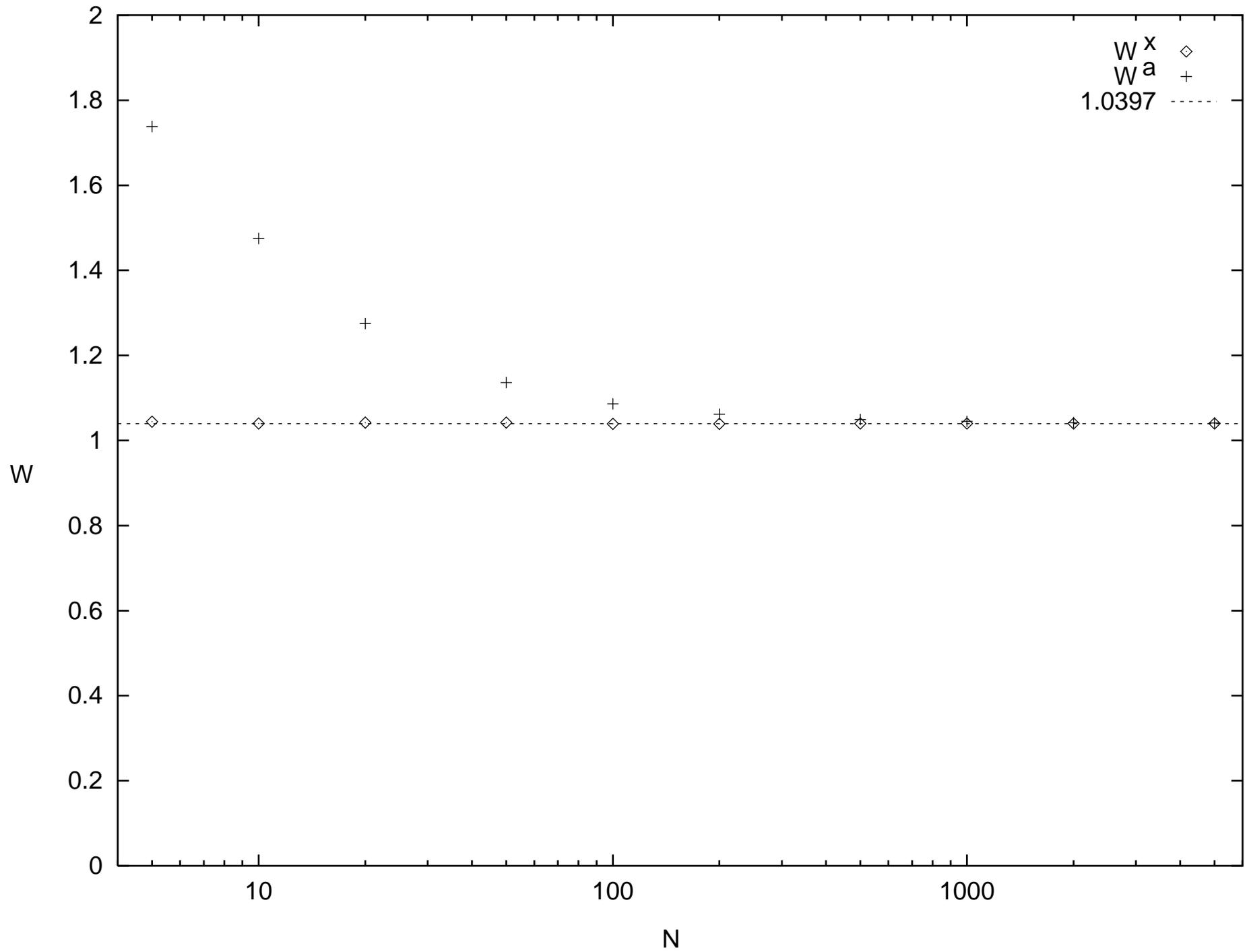
\begin{figure}
\caption{
Similar to Fig.\ref{fig:lang},
except here the evolution of the thermostatted
oscillator is implemented using Monte Carlo,
rather than Langevin evolution.
The duration of a simulation is now characterized
by the number of MC steps, $N$, rather than a
switching time $t_s$.
For each of ten values of $N$, $10^5$ simulations
were carried out.}
\label{fig:mc}
\end{figure}

In the final set of simulations involving the
harmonic oscillator, we used Monte Carlo
evolution, with the Metropolis algorithm.
Here, the duration of a simulation is characterized
by the number of MC steps, $N$, rather than by 
a switching time $t_s$.
Ten different values of $N$ were considered, 
$N = 5, 10, 20, 50, \cdots 5000$, and for each
a total of $10^5$ simulations were performed.
Fig.\ref{fig:mc} shows $W^a$ and $W^x$ for each 
value of $N$; as before, the results agree nicely 
with Eqs.\ref{eq:infty} to \ref{eq:iden}.
Fig.\ref{fig:mc_dists} shows $\rho_N(W)$ --- the 
distribution of values of $W$ obtained from the $10^5$ 
simulations ---
for each of the ten values of $N$.
Although the distributions $\rho_N$ are quite
different, the integral
$\int dW\,\rho_N(W)\exp -\beta W$ 
(i.e.\ $\exp -\beta W^x$) is independent
of $N$, as shown by the values of $W^x$ in 
Fig.\ref{fig:mc}.

As a final example, we take a system 
more complicated than a harmonic oscillator,
namely, a gas of $n_p=50$ interacting particles 
inside a piston which is taken through one
cycle of pumping.
Specifically, the particles are confined 
within a two-dimensional box with hard walls,
whose initial dimensions are $1\times 1$;
over the course of the switching process, 
one of the wall first moves inward until
the area enclosed by the box is three-quarters
of its initial value, and then back out again.
This pumping of the piston is cosinusoidal:
the $x-$ and $y-$ dimensions of the box are
given by
\begin{eqnarray}
L_x(\lambda) &=& 1.0 \\
L_y(\lambda) &=& 0.875 + 0.125 \cos (2\pi\lambda),
\end{eqnarray}
where $\lambda=t/t_s$ and
the total switching time is $t_s=10.0$.

In their interactions with one another and 
with the walls, the particles act as hard
disks of radius $R=.005$;
between collisions each particle moves freely.
(Thus, over a switching time $t_s=10.0$,
a typical particle suffers several
collisions with other particles.) 
Work is performed on the gas each time 
a particle bounces off the moving wall.

Molecular dynamics was used for the evolution, 
i.e. continuous trajectories for the particles
were computed as functions of time.
However, at each time step in the integration
of the equations of motion, a single particle
was randomly selected, and a random ``kick''  
--- a discrete change in the momentum of the
chosen particle --- was generated.
The kick was then either accepted or rejected
according to the Metropolis algorithm\cite{metro},
corresponding to a temperature $\beta^{-1}=0.5$.
This simulates coupling to a heat reservoir:
if all the walls of the box were fixed, the
gas would relax to thermal equilibrium
from any initial conditions.

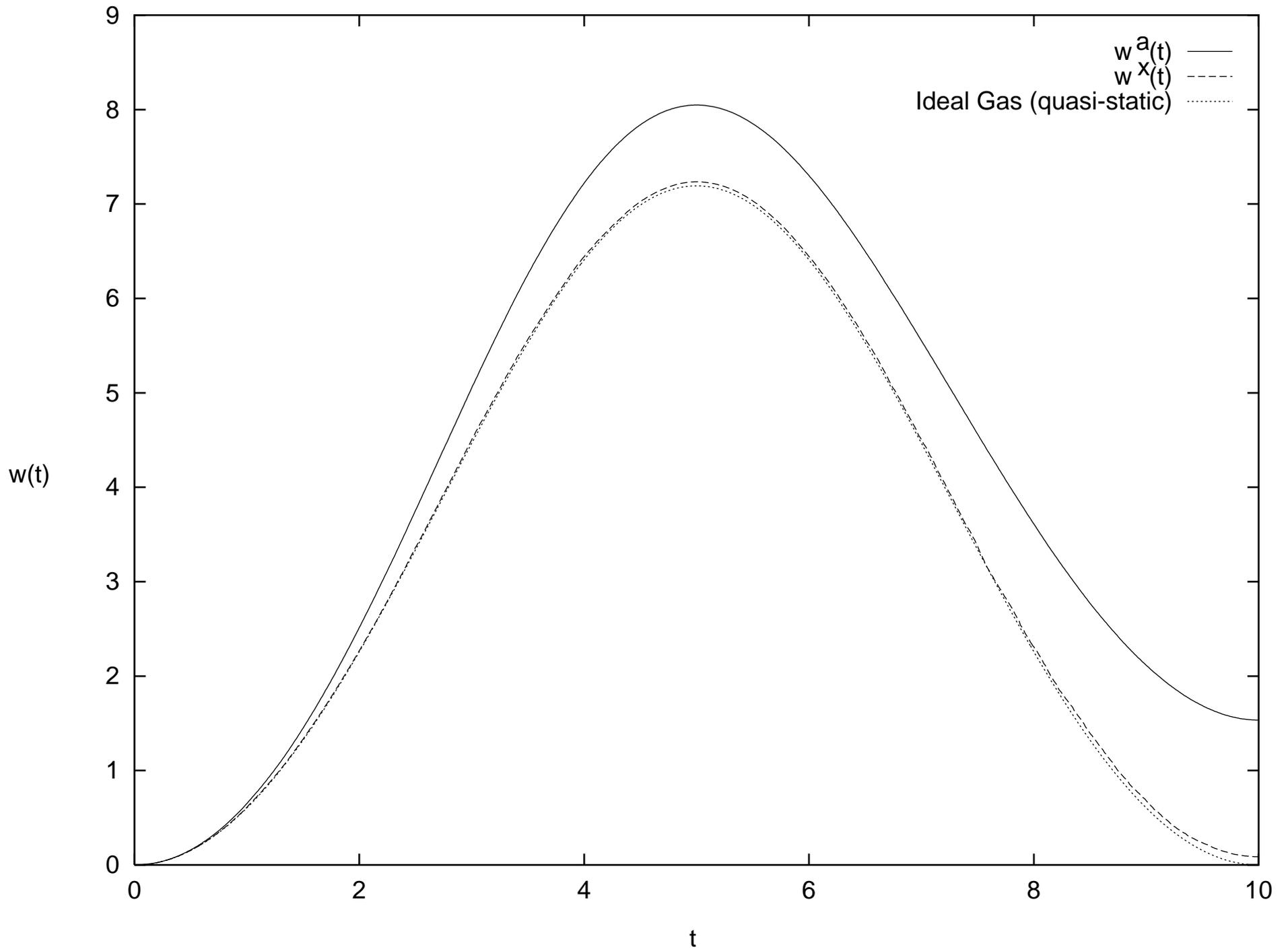
\begin{figure}
\caption{
Simulations of a gas of interacting hard disks
inside a piston which goes through one
cycle of pumping (first in, then out), over
a switching time $t_s$.
The evolution is MD, but MC ``kicks'' are
included to provide a thermostat.
A total of $10^4$ simulations were performed.
The solid line gives the average,
and the dashed line the exponential average,
of the work as a function of time,
$w^a(t)$ and $w^x(t)$, respectively.
The dotted line gives the theoretical prediction
for $F_\lambda - F_0$, Eq.\ref{eq:ideal}, 
with $\lambda=\lambda(t)$, for the case of 
an ideal gas.}
\label{fig:piston}
\end{figure}

Fig.\ref{fig:piston} shows results obtained from 
$N_s=10^4$ such simulations.
For each simulation, the initial conditions of
the gas were chosen from a canonical ensemble
($\beta^{-1}=0.5$), 
and the work performed on the gas, as a function
of time, was computed.
The horizontal axis shows time.
The solid line gives the work performed on the
gas up to time $t$ --- i.e. the ``work accumulated'',
$w(t)$ --- averaged over all $N_s$ simulations.
Let $w^a(t)\equiv(1/N_s)\sum_{i=1}^{N_s} w_i(t)$ denote
this average, where $w_i(t)$ is the work accumulated
during the $i$'th simulation.
The dashed line gives the ``exponential average''
\begin{equation}
w^x(t) \equiv
-\beta^{-1}\ln
{1\over N_s} \sum_{i=1}^{N_s}
\exp -\beta w_i(t),
\end{equation}
of the work accumulated.
Both $w^a(t)$ and $w^x(t)$ were computed from 
the same set of $10^4$ trajectories.

Since the piston returns to its initial
position at the end of the switching process,
the final free energy difference is zero:
$\Delta F \equiv F_1-F_0 = 0$.
We see that the dashed line indeed returns
to zero at $t=t_s$, with very good accuracy.
By contrast, the average work performed on 
the system (upper line) ends at 
$W^a\equiv w^a(t_s) = 1.534$.
This represents dissipated energy:
the gas ``heats up'' when pumped at a finite
rate.

At intermediate times, we expect
\begin{equation}
w^x(t) = F_\lambda - F_0 
\qquad,\qquad \lambda=\lambda(t)
\end{equation}
(in the limit of infinitely many switching
simulations),
by Eq.\ref{eq:tiden}.
If the gas were truly ideal, then the free
energy difference would be given by:
\begin{equation}
\label{eq:ideal}
F_\lambda - F_0 = 
n_p\,\beta^{-1}\ln (A_0/A_\lambda)
\qquad {\rm (ideal \,\, gas)},
\end{equation}
where $A_\lambda$ denotes the area enclosed by
the box.
However, since the particles do interact with
one another, as hard disks, this expression for
$F_\lambda-F_0$ is not exact.
Nevertheless, the size of each particle is
small enough ($R=.005$) that Eq.\ref{eq:ideal}
ought to represent an excellent approximation.
The dotted line in Fig.\ref{fig:piston} shows this 
approximation to $F_\lambda-F_0$.
This line is very close to the exponential 
average (dashed line), 
in confirmation of our central
result, in the form given by Eq.\ref{eq:tiden}.
Note that the dotted line represents the
work that {\it would have been} performed on
the gas, in the limit of infinitely slow 
switching ($t_s\rightarrow\infty$).
Thus, in the dahsed line, we have
effectively extracted this {\it quasi-static}
behavior, from an ensemble of {\it finite-time}
switching simulations!

\section{SUMMARY AND DISCUSSION}
\label{sec:sum}

The central goal of this paper has been to establish
the (exact!) validity of the result
$\overline{\exp -\beta W} = \exp -\beta\Delta F$,
within the framework of the master equation approach.
This result is unusual, in that it expresses
{\it in the form of an equality}
(rather than an inequality, e.g. Eq.\ref{eq:ineq})
the relationship between the work $W$ performed
on an out-of-equilibrium system
(more precisely, on an {\it ensemble} of systems 
driven out of equilibrium by varying an external 
parameter), and the free energy difference 
$\Delta F$ between two equilibrium states of the
system.
A few comments are now in order.

In classical statistical mechanics, the equilibrium
``state'' of a system is described by a canonical
distribution in phase space:
$f^C({\bf z}) = Z^{-1} \exp -\beta H({\bf z})$.
Its free energy is then given by
\begin{equation}
F = \langle H\rangle - \beta^{-1}S =
-\beta^{-1} \ln Z,
\end{equation}
where $\langle H\rangle = \int f^C H$
and $S = -k_B\int f^C\ln f^C$.
Thus the free energy $F$ (like the entropy
$S$) is a quantity associated with a
{\it statistical ensemble} of microscopic
states of the system.
When the system depends on some external parameter,
then so does the canonical ensemble, as in 
turn does the free energy.
The quantity $\Delta F$ of central interest
throughout this paper has been the free
energy difference (at constant temperature)
between two such equilibrium ensembles,
$f_{\lambda=0}^C({\bf z})$ and 
$f_{\lambda=1}^C({\bf z})$.

In deriving Eq.\ref{eq:iden}, we introduced a
time-dependent phase space density
$f({\bf z},t)$, describing the evolution of our
ensemble of trajectories.
Note that, although the initial phase space
density coincides with the canonical distribution
used to compute the free energy $F_0$,
\begin{equation}
f({\bf z},0) = f_{\lambda=0}^C({\bf z}) ,
\end{equation}
the final density does {\it not} (typically)
coincide with the distribution for which $F_1$ 
is defined:
\begin{equation}
f({\bf z},t_s) \ne f_{\lambda=1}^C({\bf z}).
\end{equation}
This is due to the lag which develops between the
ensemble of trajectories and an instantaneous
canonical distribution in phase space.
Thus, $\Delta F$ is not the free energy
difference between the 
initial and final states of the system\cite{not},
but rather --- as stated in the previous paragraph ---
between two different canonical
ensembles, $f_{\lambda=0}^C$ and $f_{\lambda=1}^C$,
only the former of which reflects
the actual distribution of microscopic states
of the system at any time during the switching
process.

Another way of putting this is as follows.
Suppose we are interested in the free energy
difference between two equilibrium statistical
states of a system, $A$ and $B$ (corresponding
to $f_{\lambda=0}^C$ and $f_{\lambda=1}^C$,
respectively).
Ordinarily, we would compute or measure 
$\Delta F$ by {\it reversibly} carrying the system
from $A$ to $B$, i.e. by switching $\lambda$
infinitely slowly.
Eq.\ref{eq:iden} tells us that, even if we
switch {\it irreversibly}, so that the system ends
up in some nonequilibrium statistical 
state $B^*$, we can still extract 
$\Delta F(\equiv F^B - F^A)$ from an ensemble
of such measurements.

[Of course, if we are dealing with a system which
satisfies the thermalization assumption, then, 
at the end of the switching process, we can
always, at no cost in work, hold the value of
$\lambda$ fixed and allow our ensemble to relax,
for an additional time $t_{rel}$, to thermal
equilibrium: $f({\bf z},t_s+t_{rel}) = 
f_{\lambda=1}^C({\bf z})$.
In this case the $\Delta F$ which appears on 
the right side of Eq.\ref{eq:iden} {\it does}
equal the free energy difference between the
initial and final statistical states of the
system.]

Eq.\ref{eq:iden} was derived, as an identity,
under the assumptions of Markovian evolution
and detailed balance, as spelled out in 
Section \ref{sec:prelim}.
This derivation is complementary to the one
presented in Ref.\cite{iden}, in which the
degrees of freedom of the heat reservoir were
treated explicitly.
Neither of the assumptions of Section \ref{sec:prelim}
was assumed in Ref.\cite{iden}, but the coupling
between system and reservoir was taken to be
weak, so the result there was an approximate
one, with small corrections expected from the
small but finite interaction Hamiltonian.
Of course, in a real physical system, neither
the Markov assumption nor detailed balance 
will be met exactly, so the derivation presented
herein is strictly speaking valid only for a 
particular class of {\it models} of physical
reality.
Nevertheless, because the result is exact for
these models, and because the Markov and detailed
balance assumptions are often very good 
approximations for physical systems, the result
is a useful one.
Furthermore, as illustrated in Section \ref{sec:examp},
models of thermostatted systems which are commonly
used in theoretical and numerical studies {\it do}
satisfy these assumptions; 
Eq.\ref{eq:iden} is therefore exactly valid for
these models.

It would be very interesting, of course, to find
a physical system on which a laboratory (as 
opposed to numerical!) experiment testing the
validity of Eq.\ref{eq:iden} would be feasible.
As mentioned in Ref.\cite{iden}, such a system
would almost certainly have to be micro-,
or at most meso-, scopic in size.

Finally, from both a theoretical and a
computational point of view, it would be
worthwhile to consider possible extensions or
generalizations of Eq.\ref{eq:iden}.
In particular, are analogous results valid
for ensembles other than the canonical ensemble
(fixed $N$, $V$, $T$) considered here, e.g.\
microcanonical, grand canonical, 
isothermal-isobaric, etc?
Presumably, the role of the Helmholtz free energy
$F$ would then be played by other thermodynamic
potentials, for instance the Gibbs free energy
in the case of the isothermal-isobaric ensemble.

\section*{ACKNOWLEDGMENTS}

I would like to thank J.E. Hunter III, W.P. Reinhardt, 
and J. Tams for sharing with me the results of 
unpublished numerical simulations.
These discussions have contributed significantly to 
my understanding of the practical problems likely
to arise when applying the central result of this
paper to the computation of $\Delta F$ in complicated 
systems.
This work was supported by the Polish-American
Maria Sk\l odowska-Curie Joint Fund II, under
project PAA/NSF-96-253.

\section*{APPENDIX A}

In this Appendix we present a derivation of
Eq.\ref{eq:dgdt} different from the one given
in Section \ref{sec:deriv}.

For a given stochastic trajectory ${\bf z}(t)$,
the work $W$ is given by a path integral along
that trajectory (Eq.\ref{eq:work}).
We are interested in evaluating the average
of $\exp -\beta W$ over an ensemble of trajectories,
obtained by sampling initial conditions from a
canonical ensemble and then evolving stochastically
from each of these initial conditions.
The quantity $\overline{\exp -\beta W}$ thus
constitutes a ``sum over all paths'', with
each path ${\bf z}(t)$ in our ensemble weighted 
by the factor $\exp -\beta W$.
We may write this as
\begin{equation}
\label{eq:pathsum}
\overline{\exp -\beta W} = 
\int dm\,
{\cal P}[{\bf z}(t)]\,
\exp -\beta W,
\end{equation}
where $dm$ denotes a measure in the space of paths
${\bf z}(t)$;
${\cal P}[{\bf z}(t)]$ denotes the probability
density (with respect to this measure)
of choosing ${\bf z}(t)$ by sampling
randomly from the ensemble of trajectories;
and $W=W[{\bf z}(t)]$
(as per Eq.\ref{eq:work}).

Let us now divide the time interval $[0,t_s]$
into $N$ time steps of duration
$\delta t = t_s/N$,
and let us denote a particular trajectory
${\bf z}(t)$ by its phase space locations
${\bf z}_n\equiv{\bf z}(t_n)$ at times
$t_n\equiv n\,\delta t$, 
$0\le n\le N$.
Thus,
\begin{equation}
{\bf z}(t)\rightarrow
({\bf z}_0,{\bf z}_1,\cdots,{\bf z}_N).
\end{equation}
The limit $N\rightarrow\infty$ (with $t_s$ fixed) 
is implied. 
Choosing a Euclidean measure in path space,
\begin{equation}
\int dm =
\int d{\bf z}_0\int d{\bf z}_1\cdots
\int d{\bf z}_N,
\end{equation}
the probability density for a particular path is
\begin{equation}
\label{eq:prob}
{\cal P}[{\bf z}(t)] = 
p({\bf z}_0)
P_{\lambda_1}^{\delta t}({\bf z}_0\vert{\bf z}_1)
P_{\lambda_2}^{\delta t}({\bf z}_1\vert{\bf z}_2)
\cdots
P_{\lambda_N}^{\delta t}({\bf z}_{N-1}\vert{\bf z}_N).
\end{equation}
Here, 
$p({\bf z}_0)=Z_0^{-1}\exp -\beta H_0({\bf z}_0)$
is the probability distribution for the initial 
condition ${\bf z}_0$;
$P_{\lambda}^{\delta t}({\bf z}^\prime\vert{\bf z})$
is the transition probability from ${\bf z}^\prime$
to ${\bf z}$ (in time $\delta t$) as a function of 
$\lambda$; and $\lambda_n\equiv n/N$.
It is the Markov assumption which allows this 
factorization.
The work $W$ may be expressed as
\begin{equation}
\label{eq:appwork}
W[{\bf z}(t)] = 
\sum_{n=1}^N\,
\delta H_n({\bf z}_{n-1}),
\end{equation}
where 
$\delta H_n\equiv 
H_{\lambda_n}-
H_{\lambda_{n-1}}$.
[In writing Eqs.\ref{eq:prob} and \ref{eq:appwork} we 
implicitly assume that $\lambda(t)$ evolves in $N$ 
discrete steps $\delta\lambda = 1/N$ which occur at 
times $t_0, t_1, \cdots, t_{N-1}$.
This ``staircase'' evolution becomes
$\lambda(t)=t/t_s$ in the limit $N\rightarrow\infty$.]

Combining Eqs.\ref{eq:pathsum} to \ref{eq:appwork},
we arrive at
\begin{equation}
\overline{\exp -\beta W} = 
\Biggl[
\prod_{n=0}^N \int d{\bf z}_n 
\Biggr]\,p({\bf z}_0)
e^{-\beta\delta H_1({\bf z}_0)}
P_{\lambda_1}^{\delta t}({\bf z}_0\vert{\bf z}_1)
\cdots
e^{-\beta\delta H_N({\bf z}_{N-1})}
P_{\lambda_N}^{\delta t}({\bf z}_{N-1}\vert{\bf z}_N).
\end{equation}

Let us now introduce 
\begin{equation}
g_M({\bf z}) = \Biggl[
\prod_{n=0}^{M-1} \int d{\bf z}_n
\Biggr]\,p({\bf z}_0)
e^{-\beta\delta H_1({\bf z}_0)}
P_{\lambda_1}^{\delta t}({\bf z}_0\vert{\bf z}_1)
\cdots
e^{-\beta\delta H_M({\bf z}_{M-1})}
P_{\lambda_M}^{\delta t}({\bf z}_{M-1}\vert{\bf z}),
\end{equation}
where $1\le M\le N$.
This is the discretized version of the function
$g({\bf z},t)$ introduced in the main body of the
text:
\begin{equation}
g_M({\bf z}) = g({\bf z},t_M).
\end{equation}
In particular, note that 
$\overline{\exp -\beta W} = \int d{\bf z} \,g_N({\bf z})$.
This set of functions $g_M$ satisfy the recursion relation
\begin{equation}
\label{eq:recursion}
g_{M+1}({\bf z}) = 
\int d{\bf z}_M\,g_M({\bf z}_M)
e^{-\beta\delta H_{M+1}({\bf z}_M)}
P_{\lambda_{M+1}}^{\delta t}({\bf z}_M\vert{\bf z}).
\end{equation}
Now, to first order in $\delta t$,
we have
\begin{eqnarray}
e^{-\beta\delta H_{M+1}({\bf z}_M)} &=&
1 - \beta\delta H_{M+1}({\bf z}_M) \\
P_{\lambda_{M+1}}^{\delta t}({\bf z}_M,{\bf z}) 
&=& \delta({\bf z}_M-{\bf z}) +
\delta t\,
R_{\lambda_{M+1}}({\bf z}_M,{\bf z}). 
\end{eqnarray}
Combining this with our recursion relation
gives (to leading order) 
\begin{equation}
{1\over\delta t} [g_{M+1}({\bf z})-g_M({\bf z})] = 
-\beta\,g_M({\bf z})\,
\delta H_{M+1}({\bf z})/\delta t
+ \int d{\bf z}_M\,g_M({\bf z}_M)
\,R_{\lambda_{M+1}}({\bf z}_M,{\bf z}),
\end{equation}
which becomes Eq.\ref{eq:dgdt} in the limit
$N\rightarrow\infty$.

\section*{APPENDIX B}

Here we prove the assertion (made in Section
\ref{sec:examp}) that Eq.\ref{eq:iden}
is identically true when the switching process is
carried out using the Monte Carlo method.
Some of the steps in the proof will be similar 
to those in Appendix A, but the assumption 
$N\rightarrow\infty$ will not be made here.

As mentioned in Section \ref{sec:examp}, a trajectory
$({\bf z}_0,\cdots,{\bf z}_N)$ is obtained by
alternating discrete changes in the value of $\lambda$
with random jumps in phase space generated by the 
MC algorithm.
This algorithm --- parametrized by the value of
$\lambda$ --- takes as input a point ${\bf z}$, 
and outputs a point ${\bf z}^\prime$.
Let $P_\lambda({\bf z}\vert{\bf z}^\prime)$ denote the 
probability of generating an output ${\bf z}^\prime$
from an input ${\bf z}$, for a given value of
$\lambda$.
Detailed balance is built into the algorithm:
\begin{equation}
\label{eq:mcdb}
\int d{\bf z} \,e^{-\beta H_\lambda({\bf z})}\,
P_\lambda({\bf z}\vert{\bf z}^\prime) = 
e^{-\beta H_\lambda({\bf z}^\prime)},
\end{equation}
for any $\lambda$.
(This may be accomplished by, e.g., the Metropolis
method\cite{metro}.)
Thus, a canonical distribution of inputs ${\bf z}$ gives
a canonical distribution of outputs ${\bf z}^\prime$.

The probability of obtaining a particular trajectory
$({\bf z}_0,\cdots,{\bf z}_N)$ over the course of
the entire switching process is then
\begin{equation}
\label{eq:tprob}
{\cal P}({\bf z}_0,\cdots{\bf z}_N) = 
{1\over Z_0} e^{-\beta H_0({\bf z}_0)}\,
P_{\lambda_1}({\bf z}_0\vert{\bf z}_1)\cdots
P_{\lambda_N}({\bf z}_{N-1}\vert{\bf z}_N).
\end{equation}
Combining this with Eq.\ref{eq:dwork} for the work,
we get
\begin{equation}
\label{eq:long}
\overline{\exp -\beta W} = 
\Biggl[
\prod_{n=0}^N \int d{\bf z}_n 
\Biggr]\,{1\over Z_0} e^{-\beta H_0({\bf z}_0)}
e^{-\beta\delta H_1({\bf z}_0)}
P_{\lambda_1}({\bf z}_0\vert{\bf z}_1)
\cdots
e^{-\beta\delta H_N({\bf z}_{N-1})}
P_{\lambda_N}({\bf z}_{N-1}\vert{\bf z}_N),
\end{equation}
where $\delta H_n\equiv H_{\lambda_n}-H_{\lambda_{n-1}}$.
Now notice that $\exp -\beta H_0({\bf z}_0)$ can be 
combined with $\exp -\beta\delta H_1({\bf z}_0)$
to give $\exp -\beta H_{\lambda_1}({\bf z}_0)$.
The only other factor in the integrand which depends
on ${\bf z}_0$ is $P_{\lambda_1}({\bf z}_0\vert{\bf z}_1)$.
Performing the integral $\int d{\bf z}_0$, we get
\begin{equation}
\int d{\bf z}_0 \, P({\bf z}_0\vert{\bf z}_1) \,
\exp -\beta H_{\lambda_1}({\bf z}_0) = 
\exp -\beta H_{\lambda_1}({\bf z}_1),
\end{equation}
using Eq.\ref{eq:mcdb}.
This takes care of the first of the $N+1$ integrals
appearing in Eq.\ref{eq:long}.
We now repeat this process, first combining 
$\exp -\beta H_{\lambda_1}({\bf z}_1)$
(obtained from the $d{\bf z}_0$ integration)
with $\exp -\beta\delta H_2({\bf z}_1)$
to get $\exp -\beta H_{\lambda_2}({\bf z}_1)$, 
then integrating over ${\bf z}_1$, and so forth.
At the end of this process of ``rolling up''
the factors and integrating, we are left with
\begin{equation}
\overline{\exp -\beta W} = 
\int d{\bf z}_N\,{1\over Z_0} \exp -\beta H_1({\bf z}_N)
= {Z_1\over Z_0} = \exp -\beta\Delta F.
\end{equation}
Q.E.D. 

Note that Eq.\ref{eq:goft} for $g({\bf z},t)$, derived 
within the framework of continuous-time evolution, 
also has a Monte Carlo counterpart.
Namely, for $1\le M\le N$, let us define
\begin{equation}
g_M({\bf z}) = 
\prod_{n=0}^{M-1} \int d{\bf z}_n\,
{\cal P}_M({\bf z}_0,\cdots,{\bf z}_{M-1},{\bf z})
\exp -\beta w_M,
\end{equation}
where ${\cal P}_M$ gives the probability of sampling
a particular sequence of phase space points in the first $M$
Monte Carlo steps, and $w_M$ is the work accumulated
during those steps.
Thus, in terms of our ensemble of MC trajectories,
$g_M({\bf z})$ is the weighted phase space density 
after $M$ steps, where the weight assigned to each 
trajectory is $\exp -\beta w_M$.
Then, writing an explicit expression for ${\cal P}_M$
in the form given by Eq.\ref{eq:tprob}, and rolling up
factors and integrating as above, it follows easily that
\begin{equation}
g_M({\bf z}) = {1\over Z_0}
\exp -\beta H_{\lambda_M}({\bf z})
= {Z_{\lambda_M}\over Z_0} \, f_{\lambda_M}^C({\bf z}).
\end{equation}

\begin{figure}
\caption{
For each of the ten sets of MC simulations
(see Fig.\ref{fig:mc}), the distribution of
values of work, $\rho_N(W)$ was obtained.
This figure shows these ten distributions,
from $N=5$ (lowest peak) to $N=5000$ (highest).
(Although the peak moves toward the 
right with increasing $N$, the actual average
work performed goes down; see the values of $W^a$
in Fig.\ref{fig:mc}.)}
\label{fig:mc_dists}
\end{figure}

\begin{references}

\bibitem{ll15} L.D.Landau and E.M.Lifshitz,
{\it Statistical Physics},
3rd ed., Part 1, Section 15
(Pergamon Press, Oxford, 1990).

\bibitem{ll20} {\it Ibid.}, Section 20.

\bibitem{wprjeh} 
W.P.Reinhardt and J.E.Hunter III, 
J.Chem.Phys.{\bf 97}, 1599 (1992).

\bibitem{iden} 
C.Jarzynski, Phys.Rev.Lett.{\bf 78}, 2690 (1997).

\bibitem{kubo} R.Kubo, M.Toda, and N.Hashitsume,
{\it Statistical Physics II: Nonequilibrium 
Statistical Mechanics}, Chapter 2
(Springer-Verlag, Berlin, 1985).

\bibitem{initial}
We assume these initial conditions regardless
of whether or not the thermalization assumption
mentioned above is met.
Thus even if, for instance, the system is 
isolated during the switching process, we still
assume that it was earlier allowed to thermalize 
with some heat reservoir.

\bibitem{crooks}
Manuscript in preparation.

\bibitem{ad_inv} H.Goldstein, {\it Classical
Mechanics}, 2nd ed., Sections 10-5 and 11-7
(Addison-Wesley, Reading, 1980).

\bibitem{brown} R.Brown, E.Ott, and C.Grebogi, 
J.Stat.Phys.{\bf 49}, 511 (1987).

\bibitem{hoover} W.G.Hoover, {\it Computational
Statistical Mechanics} (Elsevier, Amsterdam, 1991).

\bibitem{n} 
S.Nos\' e, J.Chem.Phys.{\bf 81}, 511 (1984). 

\bibitem{h}
W.G.Hoover, Phys.Rev.A {\bf 31}, 1695 (1985).
 
\bibitem{reviews} 
For reviews and reference books, see 
T.P.Straatsma and J.A.McCammon,
Annu.Rev.Phys.Chem.\ {\bf 43}, 407 (1992);
M.Karplus and G.A.Petsko, Nature {\bf 347}, 631 (1990);
D.L.Beveridge and F.M.DiCapua,
Annu.Rev.Biophys.Biophys.Chem.\ {\bf 18}, 431 (1989);
C.L.Brooks III, M.Karplus, and B.M.Pettitt,
Adv.Chem.Phys.\ {\bf 71}, 1 (1988);
{\it Simulations of Liquids and Solids},
D.Frenkel, I.R.McDonald, G.Ciccotti, eds., 
Chapter 2 (North-Holland, Amsterdam, 1986);
D.Frenkel and B.Smit, {\it Understanding Molecular
Simulation: From Algorithms to Applications}, 
Chapter 7 (Academic Press, Boston, 1996).

\bibitem{hph}
B.L.Holian, H.A.Posch, and W.G.Hoover,
Phys.Rev.E {\bf 47}, 3852 (1993).

\bibitem{tp}
R.Zwanzig, J.Chem.Phys.{\bf 22}, 1420 (1954).

\bibitem{bennett}
See, for instance, 
C.H.Bennett, J.Comp.Phys.{\bf 22}, 245 (1976);
G.M.Torrie and J.P.Valleau, J.Comp.Phys.{\bf 23}, 187 (1977).

\bibitem{ti}
J.G.Kirkwood, J.Chem.Phys.{\bf 3}, 300 (1935).

\bibitem{sg}
See, for instance,
T.P.Straatsma, H.J.C.Berendsen, and J.P.M.Postma,
J.Chem.Phys.{\bf 85}, 6720 (1986);
U.C.Singh {\it et al}, J.Am.Chem.Soc.{\bf 109},
1607 (1987).

\bibitem{wata}
M.Watanabe and W.P.Reinhardt, 
Phys.Rev.Lett.{\bf 65}, 3301 (1990).

\bibitem{hrd}
J.E.Hunter III, W.P.Reinhardt, and T.F.Davis,
J.Chem.Phys.{\bf 99}, 6856 (1993).

\bibitem{taiwan}
L.-W.Tsao, S.-Y.Sheu, and C.-Y.Mou, 
J.Chem.Phys.{\bf 101}, 2302 (1994).

\bibitem{chandler}
D.Chandler, {\it Introduction to Modern
Statistical Mechanics} (Oxford University,
New York, 1987), Sec.\ 5.5.

\bibitem{disease}
This is easy to see:
typically sampled values of $W$ are those 
within a standard deviation or so of the
maximum of $\rho(W)$, while the values of
$W$ which are most important in determining
$\overline{\exp -\beta W}$ are those near the
maximum of $\rho(W)\exp -\beta W$.
As a general rule, the two sets will overlap
significantly only if the function 
$\exp -\beta W$ does not change much over
one standard deviation in $W$.

\bibitem{frenkel}
Stimulating correspondence with D.Frenkel on
this point is gratefully acknowledged.

\bibitem{hh} 
W.G.Hoover and B.L.Holian,
Phys.Lett.A {\bf 211}, 253 (1996).

\bibitem{metro}
N.Metropolis {\it et al}, 
J.Chem.Phys.{\bf 21}, 1087 (1953).

\bibitem{not}
Indeed, ``the free energy of the system at 
the end of the switching process'' is not 
a well-defined quantity, since the
statistical state of the system does not correspond
to canonical equilibrium.

\end{references}
\end{document}